\documentstyle[11pt,aaspp]{article}
\def\ly{\lesssim}
\def\ga{\gtrsim}
\def\simlt{\lesssim}

\def\m{m_\nu}
\def\cdm{{\rm CDM}} 
\def\hs{\hspace{0.1cm}} 
 
\def\be{\begin{equation}}
\def\bea{\begin{eqnarray}}
\def\ee{\end{equation}}
\def\eea{\end{eqnarray}}
\def\lsb{\left[}
\def\rsb{\right]} 
\def\lb{\left(} 
\def\rb{\right)}
\def\Mi{{\rm 
Mpc^{-1}}} 
\def\M{{\rm Mpc}} 
\def\Q{{\bf Q}} 
\def\x{{\bf x}}
\def\n{{\bf n}} 
\def\k{{\bf k}} 
\def\q{{\bf q}} 
\def\e{\eta}
\def\g{\gamma}
\def\a{\alpha} 
 
\def\eb{{\bf e}}
\def\ap{a^{'}} 
\def\om{\omega} 
\def\Om{\Omega} 
\def\D{\Delta}
\def\d{\delta} 
\def\pl{\partial} 
\def\la{\lambda} 
\def\rc{\rho_{c0}}
 
\def\dq{\frac{\pl}{\pl q}}
\def\dQ{\frac{\pl}{\pl Q}} 
\def\fb{{\bar f}}
\def\ino{\int_{\e_i}^{\e}} 
\def\mp{\mu^{'}} 
\def\lp{\la^{'}}
\def\ep{\tilde \e} 
\def\F{{\cal F}} 
\def\Fh{{\hat F}} 
\def\bg{\bar g}
\def\t{\tau} 
\def\tp{{\tilde \tau}} 
\def\l{{\bf l}}
\begin{document}

\title{Decaying Neutrinos and Large Scale Structure Formation}

\author{Somnath Bharadwaj}
\affil{ Mehta Research Institute \\ Chhatnag
Road, Jhusi \\Allahabad  221 506 \\ India\\ e-mail:
somnath@mri.ernet.in}
\authoraddr{ Mehta Research Institute \\ Chhatnag
Road, Jhusi \\Allahabad  221 506 \\ India\\ e-mail: somnath@mri.ernet.in}
\and
\author{Shiv K. Sethi \altaffilmark{1}
\altaffiltext{1}{Present address: Institut d'Astrophysique, 98 bis, 
boulevard Arago, 75014 Paris, France }}
\affil{ Inter-University Centre for Astronomy \& Astrophysics \\
Post Bag 4 \\ Pune 411007 \\ India \\ e-mail: sethi@iucaa.ernet.in} 

\begin{abstract}
We study the growth of density perturbations in a universe with
unstable dark matter particles. The mass ($m_\nu$) range $30 \, {\rm
eV } \, \le m_{\nu}\, \le \, 10 \,{\rm keV}$ with lifetimes ($t_d$) in
the range $10^7 \, {\rm sec} \, \le t_d \, \le \, 10^{16} \, {\rm
sec}$ are considered. We calculate the COBE normalized matter power
spectrum for these models. We find that it is possible to construct
models consistent with observations for masses $m_{\nu}\,> \, 50 \,
\rm {eV}$ by adjusting $t_d$ so as to keep the quantity
$m^2_{\nu}({\rm keV}) t_d({\rm yr})$ constant at a value around
$100$. For $m_\nu \, \le \, 1 {\rm keV}$ the power spectrum has extra
power at small scales which could result in an early epoch of galaxy
formation. We do not find any value of $t_d$ which gives a viable
model in the mass range $m_{\nu} \, \le \, 50 \, {\rm eV}$. We also
consider the implications of radiatively decaying neutrinos---models
in which a small fraction $B \ll 1$ of neutrinos decay into
photons--which could possibly ionize the intergalactic medium (IGM) at
high redshift. We show that the parameter space of decaying particles
which satisfies the IGM observations does not give viable models of
structure formation.
\end{abstract}
\section{Introduction.}
To understand the formation of large scale structures in the universe
is one of the most challenging problem in standard big bang
cosmology. In recent years an overwhelming amount of evidence has
accumulated which suggests that the formation of structures can be
understood within the framework of the gravitational instability
picture in which the structures form through the growth of small,
scale-invariant, adiabatic perturbations created in the very early
universe by some process like inflation (see Liddle \& Lyth 1993 and
references therein). The consistency of the CMBR anisotropies measured
by the COBE-DMR (Smoot et al. 1992; Bennett et al. 1996)---which probe
the matter fluctuations at recombination--- with the present
distribution of matter on large scales supports this view and suggests
that the present universe is dominated by cold dark matter (CDM)
particles. However the standard CDM model with $\Omega_0 = 1$, $h
=.5$, and $n = 1$ normalized to COBE observations predicts neither the
correct amplitude nor the correct shape of the power spectrum of
density fluctuation at small scales (Efstathiou, Sutherland \& Maddox
1990; Efstathiou, Bond, \& White 1992; Peacock \& Dodds 1994). The
r.m.s. mass fluctuation in randomly placed spheres of radius $8
h^{-1}\, \rm Mpc$ (denoted by $\sigma_8$) provides a sensitive probe
of the power spectrum at scales around $k = 0.2 h \, \rm Mpc^{-1}$,
and various studies show that the observed abundance of rich clusters
of galaxies at the present epoch is consistent with $\sigma_8 \sim 0.5
\hbox{--} 0.8$ (Henry \& Arnaud 1991; White et al. 1993; Viana \&
Liddle 1996; Bond \& Myers 1996 Eke et al. 1996; Pen 1996; Borgani et
al. 1997; Carlberg et al. 1997), whereas the standard CDM model
normalized to the four-year COBE data predicts $\sigma_8 = 1.22$ (Bunn
\& White 1996).  In addition, the standard CDM model is also
inconsistent with the shape of the power spectrum inferred from
various galaxy surveys (Baugh \& Efstathiou 1994; Lin at. al. 1996).

Within the framework of CDM-like models, the power spectrum of density
fluctuation can be analyzed using a single parameter (Bardeen
at. al. 1986; Bond 1996) \be \Gamma \simeq \Omega_{m0} h \times \left
({\Omega_{r0} h^2 \over 4.18 \times 10^{-5}} \right
)^{-1/2}. \label{eq:i1} \ee Here $\Omega_{r0} = \rho_{r0}/\rho_{c0}$,
where $\rho_{r0}$ is the present energy density in all the
relativistic species and $\rho_{c0}$ is the present critical density,
and $\Om_{r0}$ is the contribution from the relativistic species to
the present density parameter. Similarly, $\Omega_{m0}$ is the
contribution to the present density parameter from the matter which
can `effectively' clump at small scales. While the standard CDM model
predicts $\Gamma = 0.5$, observations require that $0.22 \le \Gamma
\le 0.29$ (Peacock \& Dodds 1994), and several variants of the
standard CDM model have been proposed to overcome this
discrepancy. There are essentially two ways to decrease $\Gamma$:
1. by decreasing the amount of matter which can clump at small scales
(i.e., by decreasing $\Omega_{m0}$) or 2. by increasing the radiation
content of the universe. Models like $\lambda \rm CDM$ (Kofman et
al. 1993; Liddle et al. 1996a; Stompor et al. 1995 ), $\rm HCDM$
(Klypin et al. 1993) and $\rm oCDM$ (Liddle et al. 1996b; Yamamoto \&
Bunn 1996) all involve a decrease in the value of
$\Omega_{m0}$. Models with decaying dark matter particle achieve the
required decrease in $\Gamma$ by increasing $\Omega_{r0}$ (Bardeen,
Bond, \& Efstathiou 1987; Bond \& Efstathiou 1991; White, Gelmini, \&
Silk 1995; McNally \& Peacock 1996). In this paper we study the
decaying neutrino model in some detail.

Decaying neutrinos models are characterized by two free
parameters---$m_\nu$, the mass of the decaying neutrino and $t_d$, the
lifetime of the particle.  In the early stages of the evolution when
the temperature of the universe is much higher than $m_{\nu}$, the
energy density of the massive neutrinos is the same as the
contribution from a massless species. As the universe expands the
massive neutrinos become non-relativistic and the energy density in
this component starts increasing relative to the contribution from the
massless neutrinos. This gives rise to an era when the density of the
universe is dominated by massive neutrinos, after which these
particles decay and pump their rest energy into relativistic
particles. Once the neutrinos decay the evolution of the universe is
similar to the standard CDM model except that the energy density in
relativistic particles is higher. This extra energy in the
relativistic decay products gives the requisite increase in
$\Omega_{r0}$ (Eq. \ref{eq:i1}) which acts to lower the value of
$\Gamma$, and by suitably choosing $m_{\nu}$ and $t_d$ it is possible
to get results which are in agreement with the observed large scale
structure.

Decaying neutrino models have been investigated by earlier authors
(Bond \& Efstathiou 1991; White, Gelmini, \& Silk 1995) who find that
at the length-scales which are relevant for large scale structure
formation $(k \ly 0.1 \Mi)$ the power spectrum is essentially of a
CDM-like form with an effective $\Gamma$ parameter which has a
dependence on $m_{\nu}$ and $t_d$. This dependence captures the effect
of the enhancement of $\Om_{r0}$, and in the mass range $(m_{\nu} \ge
1 {\rm keV})$ which they have considered this is the only process
which affects the shape of the power spectrum in the relevant range of
$k$. However, if one considers neutrinos with smaller masses and
larger lifetimes,  there are other physical processes which become
important and the power spectrum starts to exhibit features which
cannot be  described by just changing the $\Gamma$ for a CDM-like
power spectrum.  For large masses the viable models are restricted to
have short lifetimes for  otherwise the energy density of the decay
product is too high and the resultant value of $\Gamma$ turns out to
be too low.  In such models the massive neutrino dominated era ends
much before any of the relevant length-scales enter the horizon. For
low neutrino masses the viable models require larger lifetimes, and
for many such models the size of the horizon in the massive neutrino
dominated era is comparable to the length-scales which are relevant
for  large scale structure formation. The main effect on the power
spectrum is that  the modes which enter the horizon in the massive
neutrino  dominated era have more power compared to a CDM-like power
spectrum and a CDM-like fit based on adjusting $\Gamma$ does not work
for these models.  

In this paper we study in some detail the power spectrum for a large
class of decaying neutrino models in the mass range  $30 \, {\rm  eV }
\,  \le m_{\nu}\, \le \, 10 \,{\rm keV}$.  In addition to the large
masses considered in earlier works  (Bond \& Efstathiou 1991; White,
Gelmini, \& Silk 1995) we have also considered models with low masses
and large lifetimes, the upper limit on the lifetime coming from the
restriction that the massive neutrino has to decay before the present
epoch.  We have not considered models where the massive neutrinos
survive until the present epoch.  While for large masses the
free-streaming of the massive neutrinos has no effect on large scale
structure formation, this process has to be taken into account when
dealing with neutrinos with low masses and large
lifetimes. Sub-horizon scale perturbations in  relativistic
collision-less particles are wiped out due to the free-streaming of
the particles. The massive neutrinos behave like relativistic
particles as long as their momentum is much larger than the rest mass,
and sub-horizon scale perturbations in the massive neutrino component
start growing  only after  the particles become non-relativistic. The
earlier works dealt with the  perturbations in the massive neutrino
component  using the hydrodynamic equations which cannot capture the
effects of free-streaming,   and hence they were restricted to large
masses. In this  paper we have used the collision-less Boltzmann
equation which  allows us to follow the free-streaming of the massive
neutrinos and hence we have been able to study models with low masses
and long lifetimes.    For a detailed discussion of free-streaming the
readers is referred to Bond \& Szalay (1983). The relevant equations
for following the growth of perturbations in a decaying massive
neutrino scenario are presented in the Appendix. 

The power spectrum for the various decaying neutrino models have been
normalized so that the theoretically predicted CMBR anisotropies are
consistent with the COBE-DMR observations. For some of the decaying
neutrino models the  normalization differs significantly from the
standard CDM model and this happens because of a large contribution
from the `integrated Sachs-Wolfe effect'.  For a discussion of the
`integrated Sachs-Wolfe effect' (Sachs \& Wolfe 1967) in the context
of some other large  scale structure formation models the reader is
referred to Kofman \& Starobinsky (1985) and  Sugiyama \& Gouda (1992).

The motivation for studying cosmological models with decaying massive
neutrinos  has been twofold. As discussed earlier, decaying neutrinos
provide possible scenarios for large scale structure
formation. Decaying neutrinos are also interesting in the context of
another very important problem in cosmology--- to understand the
ionization of the intergalactic medium (IGM) at high redshifts (Sciama
1990; Sethi 1997).  The ionization state  of the IGM at high redshifts
is inferred from Gunn-Peterson tests for neutral hydrogen, neutral
helium, and singly ionized helium, and also the proximity effect (see
eg.  Miralda-Escud\'e \& Ostriker 1990; Giroux and Shapiro 1996; Sethi
\& Nath 1996), and  it is not clear if the conventional sources of
photoionization can serve the purpose of ionizing the IGM.
Radiatively decaying neutrino models where  a small fraction  $B \ll
1$ of neutrinos decay into photons  have been proposed as a  possible
means of photoionizing the IGM (Sethi 1997).  In this paper we  have
investigated if any of the radiatively decaying neutrino models which
can explain the ionization state of the IGM can also predict a large
scale structure formation scenario which is compatible with
observations.

For the radiatively decaying neutrinos,
in addition to the two parameters $m_{\nu}$ and $t_d$,  the branching
ratio for decay into photons $B$ is another free parameter.  However,
since  $B \ll 1$ this  affects only the ionization state of the IGM.
The decay photons are dynamically unimportant   and they do not affect
the evolution of the dark matter perturbations. Thus, for the purpose
of calculating the matter power spectrum we need consider only
$m_{\nu}$ and $t_d$ as the relevant parameters and the value of $B$
is of no consequence. 

\hspace{0cm}From the point of view of particle physics models, 
we require: (1) $t_d \ll 
t_0$, $t_0$ being the present age of the universe and (2) for radiatively
decaying neutrinos, radiative lifetime ($= t_d/B$) such that the intergalactic
medium can be ionized. The latter was studied in Sethi (1997) and such models
are possible within the framework of left-right symmetric models or models
involving a charged Higgs scalar (see Fukugita and Yanagida 1995 for a recent
review). To satisfy the first condition, we need to 
construct models where all the neutrinos have decayed by the present epoch.
One of the models in which it can be  implemented is a 
non-minimal Majoran model. 
In this model the dominant decay mode of the neutrino is a massless neutrino
and a Majoran, and it is easy to get $10^{7} < t_d < 10^{16} \, \rm sec$ for
$10 \, {\rm keV} > m_\nu > 100 \, \rm eV$ (Gelmini, Nussinov \& Peccei 1992).
The models we study are quite distinct from another class of 
radiatively decaying neutrino models
suggested to solve several   galactic and extragalactic observations by 
Sciama (1990). The dominant mode of decay in Sciama's neutrinos is radiative
with radiative lifetimes of $\sim 10^{23} \, \rm sec$ with masses 
$m_\nu \simeq 30 \, \rm eV$. As these neutrinos  are essentially stable 
over the age of  the universe, structure formation with these neutrinos 
closely resembles HDM models (for more details see Sciama 1994) 
and they are not considered here. 

Finally, we briefly outline the organization of the paper. In Section
2.  we qualitatively discuss the physical effects that decide the
power spectrum in a decaying neutrino model. The detailed equations
and the computational method we use to numerically calculate the power
spectrum are presented in the Appendix. Section 3. contains the results
of our computations. We analyze the general features of the power
spectrum for different decaying neutrino models and viable models are
chosen on the basis of comparison to observations. 
In section 4. we
present the summary and a discussion of the results.   Appendix A
contains the equations which govern the evolution of
perturbations  in the metric and in the various constituents of a
decaying neutrino model. We work in the synchronous gauge
and in our analysis we treat only the scalar perturbations. 
Appendix B contains a discussion of the
initial conditions that we have used for the perturbations. Appendix C
contains the equations which we have used to calculate the CMBR
anisotropy and normalize the power spectrum. In Appendix D we briefly
describe the scheme that we have used to numerically compute the power
spectrum. 

\section{Physical effects in the decaying neutrino model.}
        In this section we qualitatively discuss some of the physical
processes  which  shape the power spectrum in a decaying neutrino
model, The detailed equations are presented in the Appendix. 

The first ingredient in any such model is the  power spectrum
of initial perturbations on scales larger than the horizon. Here we
assume that the initial perturbations  have been produced by some
viable model of inflation and we use a scale invariant
Harrison-Zel'dovich power spectrum of the form $P_i(k)=A k$. 
As the universe evolves the horizon expands, and the shape of the power
spectrum on sub-horizon scales  gets  modified by various
astrophysical processes. The 
evolution of the power spectrum in the linear regime is expressed
using the transfer function $T(k)$ which relates the final power
spectrum $P_f(k)$ to the initial power spectrum i.e. $P_f(k)=T(k)
P_i(k)$.  
Below we  discuss the various physical processes that decide the
shape of the matter transfer function in a universe with decaying
neutrinos. 

In all the models that we have studied CDM particles are the
most dominant component of the present universe and it is the
evolution of perturbations in this component which is of primary
interest. The CDM perturbations are coupled to the
other constituents of the universe through the
gravitational potential which at any epoch is largely due to the most
dominant component of the universe at that epoch. So we first discuss
which of the various components dominates the universe in the
different stages of its expansion and later discuss how this decides
the shape of the transfer function. 

In a spatially flat universe the  evolution of the scale factor is
governed by the equation  
\be
\frac{d}{d \e} a(\e)=\sqrt{\frac{8 \pi G}{3} \rho(\e) a^4 (\e)}= H_0
\sqrt{\om(\e)}\;. \label{eq:s1} 
\ee
where  $\eta(t)=\int_0^t (1/a(t^{'}))\, d t^{'}$ is the conformal time and
 $H_0 (=100 h \,{\rm km/s/Mpc})$
is the present value of the Hubble parameter, $\rho(\e)$ is the total energy
density of the universe (for decaying neutrino models it
is contributed by the photons, 2 massless neutrinos, CDM particles,
massive neutrinos, and neutrino decay product) and  $\om(\e)= \rho(\e)
a^4 (\e)/ \rho_{c0}$, 
where $\rho_{c0}$ is the present value of the critical density. 
The variable $\om(\e)$ has been chosen so that at present it coincides
with the value of the density parameter $\Om_0=1$. 

 We first consider  a universe with only two kinds of constituents  1.
relativistic particles (referred  to as radiation) which at present
contribute $\Om_{r0}$ to the density parameter, and 2. pressureless massive
particles (referred to as matter) which at present contribute
$\Om_{m0}$ to the density parameter.  The  radiation make a constant
contribution  
$\om_{r}(\e)=\Om_{r0}$  while the contribution
from matter   $\om_{m}(\e)= a(\e) \Om_{m0}$ increases as
the universe expands, and the  universe proceeds from a  radiation
dominated  era to a matter dominated era as it expands. The  
transition between the two regimes occurs when the matter 
and radiation make equal contributions to $\om(\e)$,  and  the value of
the conformal time at the epoch of matter-radiation equality is given
by  
\be
\e_{\rm eq}=2 (\sqrt{2}-1) \frac{\sqrt{\Omega_{r0}}}{\Omega_{m0} H_0} \,.
\label{eq:s2}
\ee

In the radiation dominated era the Jeans length is of the order of
the horizon ($\sim c \e$), and perturbations in all components
grow  $({\rm i.e.} \, \, \d(k,\e)\ \propto \e^2, {\rm where \hs the \hs mode
\hs} k= 2 \pi /
\lambda)$ when they  are  on scales larger than the 
horizon (i.e. they satisfy $  k c \e \ll 1$). The   growth stops if the modes
enter the horizon (i.e. $  k  \sim \pi/ (c \e)$) in the
radiation dominated era. Once the universe gets
matter dominated, matter perturbations on all 
scales grow in the same way $(\d(k,\e) \propto \e^2)$. Using
$k_{\rm eq}=\pi/(c \e_{\rm eq})$ to denote the mode which enters the horizon at
the epoch of matter-radiation equality, we can say that all modes with
$k < k_{\rm eq}$ entered the horizon in the matter dominated era and they
all grow  by the same factor during the course of the evolution. The
modes with $k>k_{\rm eq}$ enter the horizon in the radiation dominated era
and as a consequence they experience a period of stagnation when they
don't grow and the amplitude of these perturbations is suppressed by the
factor $(k_{\rm eq}/k)^2$ relative to the modes which enter in the matter
dominated era.  This fact can be used to  crudely model the transfer
function using  
the simple  form $T(k)=1$ for $k < k_{\rm eq}$ and $T(k)=(k_{\rm eq}/k)^4$
for $k \ge k_{\rm eq}$, and it depends on just one quantity---the mode which
enters the horizon at the epoch of matter radiation equality 
\be
k_{\rm eq}=\frac{\pi}{2 (\sqrt{2}-1)}\frac{\Omega_{m0}}{\sqrt{\Omega_{r0}}}
\frac{H_0}{c} \,. \label{eq:s3}
\ee

 In the standard CDM model $\Omega_{r0}$ $(=4.18 \times 10^{-5} h^{-2})$ has
contributions from the CMBR photons and 3 massless 
neutrino species, and the matter, which is largely made up of CDM
particles has  $\Omega_{\rm m0} = \Omega_{\rm CDM0}= 1$, and we have
 $k_{\rm eq}= 0.2 h^{2} \rm Mpc^{-1}$. It should be noted that here, and in the
rest of the paper, we have ignored the dynamical effect of the baryon
density $\Omega_{\rm B0}$.

The decaying neutrino model has all the ingredients of the  standard CDM
model the only difference being that one of the three neutrinos is
massive. For neutrino masses $m_{\nu} < 1 \, \rm MeV$ the massive
neutrinos decouple when they are relativistic, and after this the
comoving number density of the neutrinos remains fixed, and at present
we expect to find around 112.5 neutrinos/$\rm cm^{-3}$ 
(for the present temperature of CMBR $T_0 =  2.73 \, \rm K$).  Once the 
temperature of
the neutrinos  falls below $m_{\nu}$  the massive neutrinos 
become nonrelativistic and they contribute to the matter density,  and, if
the neutrinos remain stable up to the present epoch,
 this contribution at present  will be $\Om_{m \nu0}=m_\nu/(93.6
h^2 \, \rm eV)$. So at any
epoch before the massive neutrinos decay we have $\om_r(\e)= 3.62
\times 10^{-5} h^{-2}$ (from the photons and two massless neutrinos) and 
$\om_m(\e) = a(\e)(\Om_{\rm CDM0}+\Om_{m \nu0}) = a(\e)(1 + \Om_{m
\nu0})$, and the epoch of matter radiation equality is at $t_{\rm
eq1}$ 
with 
\be 
\e_{\rm eq1}=2 (\sqrt{2}-1) \frac{\sqrt{\Omega_{r0}}}{(1+\Omega_{m \nu 0})
H_0} \,.     \label{eq:s4}
\ee
The universe remains matter dominated from the epoch $t_{\rm eq1}$ to
$t_d$ when the massive neutrinos decay into 2 massless particles
referred to as the relativistic decay product. When the massive
neutrino decays, the matter
density falls from  $\om_m(\e)=  a(\e)(1 + \Om_{m\nu0})$  to
$\om_m(\e)=  a(\e)$, and the rest energy 
of the massive neutrinos gets converted into radiation. This causes
$\om_{r}(\e)$ to increase from $\Om_{r0}$ to $\Om_{r0}+ a_{\rm d}
\Om_{m\nu0}$  (where $a_d=a(t_d)$) and the universe  becomes
radiation dominated for a second time. The second 
radiation dominated era starts at 
\be 
\e_{d}= \frac{1}{H_0} \lsb \frac{12 t_d H_0}{(1+\Om_{m \nu0})} \rsb^{1/3}
\label{eq:s5}
\ee
when 
\be
a_{d}=\frac{(1+\Om_{m \nu 0})^{1/3}}{4} \lsb 12 t_d H_0 \rsb^{2/3}
\label{eq:s6}
\ee
and after this the evolution  is just like the standard CDM model except that
in this case the radiation density is higher. The universe has a
second epoch of matter-radiation equality at
\be 
\e_{\rm eq2}=2 (\sqrt{2}-1) \frac{\sqrt{\Omega_{r0}+a_d \Om_{m\nu0}}}
{H_0} \,. \label{eq:s7}
\ee
and after this the universe is dominated by the CDM particles.
The shape of the transfer function is determined by the modes which
are entering the horizon at the epochs when the universe changes from
radiation to matter dominated and vice-versa. Expressing these in
terms of the corresponding mode in the standard CDM model
(i.e. $k_{\rm eq} = 0.2 h^2 \, \rm Mpc^{-1})$ , we have  
\bea
 k_{\rm eq1}&=& 1.07 (1+\Om_{m \nu0}) k_{\rm eq} \, , \\\label{eq:s8}
 k_{d}&=&\lsb  \frac{3.97 \times 10^9 \,
 {\rm sec}}{t_d} (1+\Om_{m \nu0}) \rsb ^{1/3} 
h^{-4/3} k_{\rm eq} \, , \\ \label{eq:s9}
\eea
and
\be
k_{\rm eq2}={k_{\rm eq}}\times \left [{{0.866 + \lb 
\frac{t_d}{5.59 \times 10^{10}{\rm sec}} 
\rb^{2/3} (1+\Om_{m \nu0})^{1/3} \Om_{m \nu0} 
h^{8/3}}}\right ] ^{-1/2}\,.\label{eq:s10}
\ee 

 For the modes which enter the horizon after the neutrinos decay the
transfer function  is of the same form as the standard CDM transfer
function with $T(k)=1$ for $k< k_{\rm eq2}$, and $T(k)=(k_{\rm eq2}/k)^4$ for
$(k_{\rm eq2} \le k  < k_{d})$. Also, as  $k_{\rm eq2} < k_{\rm eq}$, the power
is reduced on larger scales as compared to the CDM model. 

 All the modes which enter the horizon in the first matter
dominated era grow by the same factor and they experience a
period of stagnation when the universe is radiation dominated for
the second time. The amplitude of these modes is suppressed by the
factor $(\e_{\rm d}/\e_{\rm eq2})^2$ relative to the modes which enter in the
final matter dominated era, and the transfer function has the form
$T(k) = (k_{\rm eq2}/k_{d})^4$ in the range $(k_{ d} \le k <
k_{\rm eq1})$. The modes which enter the horizon in the first radiation
dominated era experience a further period of stagnation corresponding
to the interval between their entering horizon and $\e_{\rm eq1}$
when the universe becomes matter dominated for the first time, and
for $k \ge k_{\rm eq1}$ the transfer function has the form 
$T(k) = (k_{\rm eq2}/k_{d})^4 (k_{\rm eq1}/k)^4 $. 

In figure 1 we show  the contribution to $\om(\e)$ from the various
species for a universe with $h= 0.5$,  and a decaying neutrino with
$m_{\nu}=200 \, \rm  eV$ and  $t_d=10^{13} \, \rm s$.  In this figure the 
contribution to $\om(\e)$ from the different constituents is plotted
against  the mode $k=\pi/c \e$ which is entering the horizon at that 
epoch. This illustrates how the dominant component changes and
which are the modes inside the horizon as these changes occur. 

In figure 2 we show the transfer function for this decaying neutrino
model based on the simple considerations discussed above. In the same
figure  we have also shown the results of a more detailed numerical
computation of the transfer function for the same decaying neutrino
model.  In the latter calculation we have numerically  followed the
evolution of both the metric, and the relevant properties of the
different constituents. For the background universe we have solved for
the scale factor along with the density of the different constituents.
For the  massive neutrinos we have used the distribution function 
to calculate  the background density and  this accurately follows the
evolution of these particles from the relativistic 
to the nonrelativistic regime. This also takes into account the
fact that relativistic time dilation causes the faster moving neutrinos
to live longer in the frame of the cosmological observer.  

We have calculated the evolution of the metric perturbations in the
synchronous gauge and we have dealt with only the scalar part of the
perturbations. For the CDM perturbations we have used the pressureless
hydrodynamic equations. We have treated the photon-baryon fluid in the
tightly coupled limit and we have ignored the baryon density in studying
the dynamics. For the massless neutrinos, the massive neutrinos and
the relativistic decay product we have used the collisionless
Boltzmann equation to follow the evolution of perturbations of the
distribution 
function, and we have used this to calculate the perturbation in the
density, pressure and the anisotropic stresses. This takes into
account the fact that the relativistic neutrinos have a very large
mean free path (of the size of the horizon) and all sub-horizon
perturbations in this component are wiped out due to the
free-streaming of these neutrinos. Sub-horizon
perturbations in the massive neutrino component grow only after the
neutrino becomes nonrelativistic. The evolution of perturbations in the
other components affects the CDM perturbations only through the metric
perturbation and the Appendix contains a more detailed description of
the equations we have used to follow the evolution of the
perturbations. 

In figure 2 we also show the numerically computed transfer function
for the standard CDM model with $h=0.5$. We see that the  main effect
of the decaying neutrinos is to reduce power on all scales except on a
small range of scales near the first matter radiation equality $k_{\rm eq1}$.

We have normalized the power spectrum for the decaying neutrino models
by calculating the r.m.s. quadrupole moment $Q_{\rm rms}$ of the  angular 
distribution of the temperature fluctuation in the CMBR, and we have
normalized the power spectrum to $Q_{\rm rms}=17 \mu \rm K$. This is
consistent with data from four years of COBE-DMR observations  (Bunn \&
White 1996).  
  On large angular scales $(\theta > 1^\circ)$  
the major contribution to the anisotropy in the CMBR is due to the
linear  Sachs-Wolfe effect (Sachs \& Wolfe, 1967) which is an integral
of the $\e$ derivative of the  metric perturbations along the photon's
trajectory from the last scattering surface to us. In the standard CDM
model the photons 
decouple from the baryons in the matter dominated era, and this
integral  is restricted to the matter dominated era and it can be
reduced to a surface term which 
relates the anisotropy in the CMBR to the fluctuations in the
gravitational potential at the last scattering surface. 
In the decaying neutrino models  we have two radiation dominated era,
and there may be significant contributions to the CMBR anisotropies
from the metric perturbations in the second radiation dominated
era. For these models it is not possible to analytically calculate the
CMBR anisotropy, and we have numerically evaluated the integral over the
metric perturbations to calculate the CMBR anisotropy. We have used
this to normalize the power spectrum for  all the decaying neutrino
models and  the details are presented in the Appendix. 

The normalization is significantly different from the standard CDM
model in only those cases where the second matter-radiation equality
occurs quite late i.e. ($\e_{\rm eq2} \sim \e_0$, where $\e_0$ is the
value of the conformal time at present). The extreme case of such a
situation is where the universe remains radiation dominated
until the present epoch and such cases can be ruled out from age
constraints. If the universe remained radiation dominated until the
present epoch,  the age of the  universe would be  $1/2 H_0^{-1}$ as
compared to  $2/3 H_0^{-1}$ for the  matter  dominated case, and an
age as short as  $\sim 10^{10} \, \rm Gyr$ is ruled out by the
observations of the oldest  globular clusters. We have only
considered those models where the universe is matter dominated at
present. As we shall see in the next section, there are models
which satisfy this criterion but still have a significant contribution
from the  `Integrated Sachs-Wolfe effect,' and the  normalization of
the power spectrum for these models is quite different from the
standard CDM normalization.  

It is possible to think of  the HCDM models as a limiting case of a
decaying neutrino model with a  small neutrino mass and 
$t_d \gg t_0$, however, there is a qualitative difference between the HCDM
models and the decaying neutrino models. For the decaying neutrino
models the constraint that the age of the universe is required to be
$\simeq 2/3 H_0^{-1}$  implies that the energy in the form of the
relativistic  decay products must be negligible at  present, and 
the universe is largely dominated by the CDM particles ($\Omega_{\rm
CDM} \sim 1$) which clump at small scales. This should be contrasted
with the the HCDM models where $20\hbox{--}30 \%$ of the matter 
at present is in the form of low mass  neutrinos which do not clump at
small scales.

\section{Results}
In this section we analyze the  numerically computed power spectrum
for a large class of decaying neutrino models and we compare our results  
with observations at various length scales. 

  Allowed models are selected on the basis of comparison to:
\begin{itemize}
\item[1.] the observed cluster abundance (which probes scales $\simeq
8  h^{-1} \, \rm Mpc$) (White et. al 1993; Eke et. al 1996;
 Henry \& Arnaud 1991;
Viana \& Liddle 1996;  Bond \& Myers 1996;  Pen 1996; Borgani et
al. 1997; Carlberg et al. 1997),
\item[2.] the peculiar velocity measurements on scales up to $60 h^{-1}
\, \rm Mpc$  (Bertschinger et al. 1990; Courteau et al. 1993; 
for comprehensive reviews see Strauss \& Willick  1995; Dekel 1994 and 
references therein ),
\item[3.] the three-dimensional power spectrum derived from the APM
survey (Baugh \& Efstathiou 1994),  which determines the shape of the
power spectrum on scales from $1  h^{-1} \, \rm Mpc$ to $ 50h^{-1} \,
\rm Mpc$, and  
\item[4.] the power spectrum obtained from the
Las Campanas Redshift Survey (LCRS) (Lin et al. 1996). 

We also compare the predictions of the allowed decaying neutrino
models with  the  observed abundance of damped Lyman-$\alpha$ clouds
at high redshifts.

\end{itemize}
\subsection{Analysis of the linear power spectra.}
For the  scales which enter the horizon after the massive neutrinos decay
(i.e. $k<k_d$), the massive neutrino model is akin to the standard CDM
model, the only difference being the delay  in the matter-radiation
equality due to the additional radiation density contributed by the
decay products.  The delay in matter-radiation equality can be cast as
a change in the `shape parameter' $\Gamma=5 \,  \hs k_{\rm eq}
h^{-1} \, \rm {\rm Mpc}$  for the power spectrum in CDM-like models,
and for the decaying neutrino models we can similarly  use
$\Gamma= 5  \hs  \, k_{\rm eq2} h^{-1} \, {\rm Mpc}$, where  $k_{\rm
eq2}$    can be calculated using equation (\ref{eq:s10})
This definition of $\Gamma$ is the same as the one used in equation 
(\ref{eq:i1}) when the value of $k_{\rm eq}$ is expressed in terms of
$\Om_{r0}$ and $\Om_{m0}$. A point to note is that in the decaying
neutrino models $\Gamma$ depends on the parameters $m_{\nu}$ and $t_d$
through the combination $t^2_d (1+m_{\nu}/(93.6 h^2 \, {\rm eV}))
m^3_{\nu}$   (eq. \ref{eq:s10}). and there is
a degeneracy in the relation between  $m_{\nu}$, $t_d$ and $\Gamma$.  
In the limit $m_{\nu} \gg 93.6 h^2 \, \rm eV$ this relation is simpler
and we find $\Gamma$ depends on just the combination $m^2_{\nu} t_d$,  
which  is in agreement with  White, Gelmini, \& Silk (1995), and
contrary to the $m_{\nu} t_d$ scaling obtained by Bond \& Efstathiou
(1991). 
  
A  comparison of various numerically computed power spectra with  the fit
using $\Gamma=5 \hs k_{\rm eq2} h^{-1} \, \rm Mpc$ 
shows that for large masses this
slightly underestimates the value of $\Gamma$, and we obtain a better
fit using  a minor variant of the  form for $\Gamma$ proposed by
White, Gelmini, \& Silk (1995). 

We find that for  large masses $m_\nu \ga 1 \, \rm keV$ and small
lifetimes $t_d \le 100 \rm yr$,  the power spectrum in the range $k< 1
h^{-1} \rm Mpc^{-1}$ is well described by a CDM-like power spectrum 
\begin{equation}
P(k) = A k \times \left[1+(ak +(bk)^{3/2} +(ck)^2)^\nu \right]^{-2/\nu},
\end{equation}
 with $a = (6.4/\Gamma)  h^{-1} \, \rm Mpc$, 
$b = (3/\Gamma)  h^{-1} \, \rm Mpc$, and 
$c = (1.7/\Gamma) h^{-1} \, \rm Mpc$. 
where our best fit $\Gamma$ is:
\begin{equation}
\Gamma \simeq  h \times \left( 1+ 0.15 \left[\left({m_\nu \over 
1 \, {\rm keV}}\right)^2
{t_d \over 1\, {\rm yr}}\right]^{2/3}\right)^{-1/2}, \label{eq:r1}
\end{equation}
which differs  from the fit given by White, Gelmini, and Silk (1995)
only in a small  change of the  numerical coefficient.
The value of normalization $A$ is determined from COBE observations
and may contain a substantial contribution from the `integrated Sachs-Wolfe' 
effect (Section (2) and Appendix~C). For these decaying neutrino
models  ($m_\nu \ga 1 \, \rm keV$ and $t_d \le 100 \rm yr$) a pure
CDM-like fitting function correctly  describes the  linear power spectrum
for the purposes of comparing the  theoretical predictions with
observations like peculiar velocities of galaxies, galaxy-galaxy
correlation functions and cluster  abundances which probe the power
spectrum at scales $k< 1 h \Mi$. At smaller scales ($k> 1 h \Mi$) these
models start exhibiting an enhancement of power      
due to the  matter dominated era caused by the  massive neutrinos
before they decay,  and a  CDM-like fit is inappropriate. 
 
For smaller neutrino masses $m_\nu < 1 \, \rm keV$ and long
lifetimes $t_d \gg 100 \rm yr$, the mode $k_d$ enters the range $k \le
1 h \Mi$a and a  CDM-like fit  is not adequate to describe all  the
features of the power spectrum at these length-scales.  We find 
that a CDM-like form correctly describes the initial deviation (at
$k \sim k_{\rm eq2}$) of the power spectrum from the primordial Harrison
Zel'dovich form, but it fails  on scales $\sim k_{d}$ where
there is a significant enhancement of power in the decaying neutrino
models. The enhancement of power is most pronounced at the scales
$k_d \ly k \ly k_{eq1}$ which entered the horizon  when the
universe was dominated by the massive neutrinos. The scales
$k>k_{eq1}$ entered the horizon during the era when the massive
neutrino was relativistic and the universe was experiencing the first
radiation dominated phase. The growth of  perturbations on these
scales is suppressed due to free-streaming of the neutrinos and the
oscillations in the photon-baryon fluid.    

In figure \ref{fig:all} we show several power spectra for
different decaying 
neutrino models  with $h=0.5$, and $m_\nu$ and $t_d$ such that for all of
them the product $m_{\nu}^2 t_d$ has the  constant value  which,
according to equation (\ref{eq:r1}),  corresponds to  $\Gamma \simeq
0.24$.  In the range $k \le 5 h \Mi$ the curve for $m_{\nu}= 10 \rm
keV$ is indistinguishable from  a CDM-like spectrum with
$\Gamma=.24$. As $m_{\nu}$ is reduced the power spectrum starts
deviating from the pure CDM-like form and it starts
exhibiting the various small scale features mentioned above and
discussed in detail in section 2. The change in small scale power in
the linear power spectrum can be quantified using $\sigma_R$ which is
the  theoretically predicted r.m.s. mass fluctuation in randomly
placed spheres of radius $R h^{-1} \M$. It should be noted that the
$\sigma_R$ discussed here is calculated using linear theory and for
scales smaller than $R=8$ non-linear effects have to been taken into
account before this  can be compared with observations. The change in
small scale power as the neutrino mass is decreased is shown for
various length-scales in figure \ref{fig:sig},  the largest scale
being $8 h^{-1} \M$ where the mass fluctuation is observationally very well
determined and the smallest being $.25  h^{-1} \M$ which is comparable
to the scales important for the formation of  damped Lyman-$\alpha$
clouds discussed later in this section. We find that as the mass is
decreased there is nearly a two-fold increase in the small scale
power compared to what  one expect from a pure $\Gamma=.24$ CDM spectrum. 
At the scale $8  h^{-1} \M$ the mass fluctuation keeps on increasing
monotonically as $m_{\nu}$ is reduced and this results in some of
the low mass models  predicting too high a value of $\sigma_8$ to be 
compatible with observations. At smaller scales $\ly 2 h^{-1} \M$ 
the mass fluctuation initially increases as $m_{\nu}$ is reduced but
then starts decreasing again. This happens because for low neutrino
masses these scales enter the horizon before the massive neutrinos
become non-relativistic and the growth of fluctuations at these 
scales is suppressed.

Figure \ref{fig:all} also shows  the APM power spectrum with a  bias
parameter $b 
= 1.2$. Also shown is the best fit to LCRS  power spectrum  (Eq. 23 in Lin et
al.'s paper)  with $b = 1.2$. While there are uncertainties in the 
value of the bias parameter for the galaxy surveys, it is believed
that the value of $\sigma_8$ as determined from cluster abundances is  
relatively free of bias. We have chosen  the value $b=1.2$ for
the APM survey to make the observed power spectrum compatible with 
$\sigma_8 =0.55$, and we have used this to visually compare the shape
of observed power spectrum with the theoretically predicted power
spectra for various decaying neutrino models. For comparison we have
also shown the $\Gamma=0.5$ CDM power spectrum  which is obtained in
the standard CDM model with $h=0.5$. While we have shown the linear power
spectrum for the various theoretical models, the observationally
determined power spectrum has significant  non-linear effects at $k >
0.2h \Mi$ and a detailed comparison is not possible at these scales
until these effects have been taken into account.

We next briefly discuss the COBE normalization of the power spectrum
for the different models. All the models shown in figure
\ref{fig:all}  have the same normalization which matches with the CDM
normalization, and this happens because for all these models the second
radiation dominated era ends much before the present epoch i.e. 
$(\eta_{\rm eq2} \sim .01 \eta_0)$. In
figure \ref{fig:nal} we show the  power spectrum for the decaying
neutrino model with $m_{\nu}=100 \,  \rm eV, t_d=5 \times  10^{14} \rm sec$ for
which the second radiation dominated phase ends much later $(\eta_{\rm
eq2} \sim .1 \eta_0)$ and this  corresponds to $\Gamma 
\sim .02$. We find that for this model the normalization is significantly
lower than the CDM normalization due to the contribution from the
`integrated Sachs-Wolfe effect'. We have considered a large number of
different models where  $\eta_{\rm eq2} \sim .1  \eta_0$, all of
which satisfy the constraint that $t_0 \sim (2/3) H_0$ and we find
that in all of them the effect is similar to the case shown in figure 
\ref{fig:nal}. 

\subsection{Allowed Models}
A very  sensitive observation for fixing the value of $\sigma_8$ is the 
abundance of rich clusters at the present epoch (Evrard 1989). Taking 
into account various
uncertainties due to cluster temperatures, cluster x-ray luminosities, and the 
 N-body methods in comparing theoretical estimates with observations
fix the value of $\sigma_8$ between $0.5$
and $0.9$ (White et. al 1993; Eke et. al 1996; Henry \& Arnaud 1991;
Viana \& Liddle 1996;  Bond \& Myers 1996;  Pen 1996; Borgani et
al. 1997; Carlberg et al. 1997). Larger values  also seem to be ruled
out by constraints from the pairwise random motions of  galactic size
dark matter halos (Gelb and Bertschinger 1994). 

For a decaying neutrino model to be consistent with the galaxy
surveys (APM, Baugh \& Efstathiou 1994; LCRS, Lin et al. 1996) at large
scales we have used the broad  criterion that equation (\ref{eq:r1}) should
predict a value of $\Gamma$ in the range $.2 \le \Gamma \le .3$ for the
model,  and we have discussed a set of models for which $\Gamma=.24$
in some detail.    

 In Table~1 we list the value of $\sigma_8$ for a wide range of
decaying neutrino models. 
A class of models allowed by this observation is  $m_\nu = 1 \, \rm
keV$ and  $t_d = 100 \, \rm yr$,  and all the models obtained from
this model keeping $m_\nu^2 t_d$ constant for  $m_\nu > 50 \, \rm
eV$ (figure \ref{fig:sig}), These models correspond to
$\Gamma=0.24$ at large scales.  Not only do these models 
produce a value of $\sigma_8$ in the right range, but as seen in
figure \ref{fig:all}, they also reproduce the shape of the observed
galaxy power spectrum quite well. A comparison of the observed bulk
velocities at  scales $\sim 40 \, \rm Mpc$ and  
$\sim 60 \, \rm Mpc$  with the predictions of these  models show that
these models may not have enough large scale power (figure \ref{fig:vel}). 
However, due to
systematic uncertainties in the analysis of the data and statistical
uncertainties due to cosmic variance,  the actual peculiar velocities
can be much lower than the values  shown in (figure \ref{fig:vel})
(Dekel 1994),  and therefore a marginal  inconsistency of our result with
peculiar velocity measurements cannot rule out these models. 

For the above class of models (i.e. $(m_{\nu}/1{\rm keV})^2 
(t_d/1 \rm yr) = 100 $) the value of $\sigma_8$ keeps on increasing as
the mass is lowered below $\sim 200 \rm eV$ and and the mass range
$m_\nu \le 50 \, \rm eV$  predicts too large a value of $\sigma_8$ to be
compatible with  observations of cluster abundances. Models with small
masses  ($m_\nu \le 50 \, \rm eV$) are ruled out by the cluster
abundance constraints irrespective of the value of $t_d$. We have run
models for $m_\nu \le 50 \, \rm eV$ with $t_d$ varying over seven 
decades from $10^{9}\, \rm sec$ to $ 10^{16} \, \rm sec$
and  and we do not find any for which $\sigma_8 \le 0.9$
The upper limit on the value of 
$t_d$ is chosen from the consideration of keeping the age of the universe 
$\sim 2/3 H_0^{-1}$.  The reason why it is not possible to get
acceptable models in the low mass range is easy to understand. For the
decay products to substantially delay the final matter radiation
equality the massive neutrinos have to decay after they have become
non-relativistic and they have a significant amount of rest energy to
pump into the decay products. For this to happen these
neutrinos should decay only after they have caused the universe to  become
matter dominated and remain matter dominated for some time. The epoch
when the first matter radiation equality ($\eta_{\rm eq1}$) occurs
depends only on the mass of the neutrino (equation \ref{eq:s4}) and
for masses 
$m_{\nu} \le 50 \rm eV$ the mode $k_{\rm eq1}$ is very close
to the mode $0.2 h \Mi$ which is  the scale where the power
spectrum is probed by $\sigma_8$. As discussed in section 2, the power
in the range of modes $k_{\rm eq1} \ge k \ge k_{\rm d}$ is enhanced as
a consequence of the first matter dominated era, and this gives rise
to a large value of $\sigma_8$.  Since $k_{\rm eq1}$ has no
dependence on $t_d$, the value of $\sigma_8$ too is nearly independent
of $t_d$ for these models. 

        We find that by varying both $m_{\nu}$ and $t_d$ it is
possible to construct a large number of models all of which predict
values of $\sigma_8$ which are in the correct range, and some of these
are shown in Table~1 and figure \ref{fig:nal}. 
 As one notices, models which give  power spectra of very different
shapes  at larger scales can still produce  acceptable values of
$\sigma_8$. However, most of these  models have too small a power at
large scales, and they do not predict a reasonable value for
$\Gamma$. In addition, the peculiar velocities predicted by these
models are too low to be consistent with observations (figure
\ref{fig:vel}).  

Going back to the allowed models, we find that an interesting region
in the parameter space  lies in the mass range $< 2 \, \rm keV$
which, apart from being in consonance with all the observations, can
give extra power at small scales (figure \ref{fig:sig}; see also,
McNally \& Peacock 1996).  This will result in an early epoch
of galaxy formation which might  ionize the hydrogen in the
intergalactic medium,  which is seen to be highly ionized up to $z
\simeq 5$ (Giallongo et al. 1994). Recent  observation  of  high
redshift galaxies  in the Hubble Deep Field (HDF) indicate that the
number density of  galaxies at $z \simeq 3$ may be comparable to the
$z \simeq 0$ population (Steidel et al. 1996; Madau et al. 1996).  
Such observations require a high redshift of galaxy formation which occurs
naturally within the framework of these decaying neutrino models.

Another possibility with the decaying neutrino scenario is to consider
radiatively decaying neutrinos. In these models, a small fraction $B$
of massive neutrinos 
decay into a photon and a massless neutrino. It is then possible
to directly ionize the intergalactic medium at high
redshifts by the decay photons (rather than by formation of high
redshift objects).  There exist a number of cosmological and 
astrophysical constraints on models of radiatively decaying
neutrinos. These include  constraints from the spectrum of the CMBR,
the supernova 1987A, the cooling of red giants, and the diffuse
extra-galactic  background of photons,  and for a detailed discussion of
constraints on radiatively decaying  neutrinos the reader is referred
to  Kolb \& Turner (1990) and references therein. The parameter space
for radiatively decaying decaying neutrinos which is allowed by all
these observations and which is consistent with the observed
ionization state of the IGM had been studied by Sethi (1997). 
where it was shown that the acceptable values of $B$ lie 
between $10^{-5}$ and $10^{-7}$. For this range of $B$, the allowed 
values of $t_d$ and $m_\nu$ lie between: 
\be
t_d \simeq 3 \times 10^{14} {\rm sec} \, (m_\nu/100 \, \rm eV)^{-2.5} 
\label{eq:ra}
\ee
\noindent and
\be
t_d \simeq 3 \times 10^{15}  {\rm sec} \, (m_\nu/100 \, \rm eV)^{-2.5}.
\label{eq:ra1}
\ee
We have calculated the power spectrum for  several of the models
allowed by the intergalactic medium ionization and in figure
\ref{fig:dec} we show 
some of the power spectra.  The  corresponding 
values of $\sigma_8$ and bulk velocities are shown in Table~1. It is
clearly seen  that all these model seem to be at variance with the
observed galaxy  power spectrum and peculiar velocity
measurements. Figure \ref{fig:reg} shows the approximate region of the
$m_{\nu}-t_d$ parameter space allowed by the IGM constraints 
(Eq. (\ref{eq:ra}) and (\ref{eq:ra1})) as compared 
to the region of  parameter space allowed by observations of the
large scale structure in the universe, and it is clearly seen that there
is no overlap between the two regions. For a given mass $m_\nu$ the
lifetimes acceptable for large scale structure formation are too short
as far as ionizing the IGM is concerned, and as seen 
in figure \ref{fig:reg}, these radiatively decaying models are ruled
out by the 
shape of the  observed power spectrum.  However, it is worth
mentioning that as noted in  Sethi (1997), several  of these models
give acceptable value for $\sigma_8$ although they predict an
unacceptable shape for the power spectrum. 
It should be mentioned that the gap
between the ranges allowed by the  structure formation and the IGM ionization
is too large to be bridged by a change in cosmological parameters. However,
the behaviour of the two allowed regions as cosmological
parameters are varied  can be qualitatively understood:
for instance if the value of $h$ is increased this would, for a given
 $\Omega_B$, increase $\tau_{\rm \scriptscriptstyle GP}$ because 
$\tau_{\rm \scriptscriptstyle GP} \propto h^3$ (see e.g., Miralda-Escud\'e
\& Ostriker 1990),  which means more ionizing photons
would be required to satisfy the GP tests. If one keeps the value of $B$, the 
branching ratio, fixed, even larger values of \{$m_\nu$,$t_d$\} are needed to
ionize the IGM (Sethi 1997). An increase in the value of $h$ would also mean
more power at small scales (an increase in the value of $\Gamma$), which will
have to compensated by an increase in \{$m_\nu$,$t_d$\}. Therefore, the 
net results of an increase in the value of $h$  would be to shift up both the 
allowed regions in figure \ref{fig:reg}. Such an effect is difficult
to quantify over an   entire range of \{$m_\nu$,$t_d$\} because of
complicated dependence of the  ionizing flux on $m_\nu$ and $t_d$. 

\subsection{High Redshift Objects.}
The  abundance of  high redshift objects like damped
Lyman-$\alpha$ clouds is an important diagnostic tool for studying
structure formation models  (Mo \&  Miralda Escud\'e 1994; Padmanabhan  \&
Subramanian 1994). The neutral hydrogen column density of damped
Lyman-$\alpha$ clouds is comparable  
to the present day spiral galaxies. Also the inferred sizes of these
objects  suggest that the damped Lyman-$\alpha$ clouds are progenitors
of the present day spiral galaxies (Wolfe at al. 1992). Observations
of damped Lyman-$\alpha$ clouds show that a large fraction of baryons
in the  form of neutral hydrogen  may already have collapsed into
forming these systems at $z \simeq 3$.  
According to  Lanzetta, Wolfe, and Turnshek (1995) the quantity
$\Omega_{\rm HI}$, which is the density of  neutral hydrogen in damped
Lyman-$\alpha$ clouds expressed as a fraction of the critical density,
can be fit by a simple relation in the redshift range  $z \simeq 0$ to
$z \simeq 3$ (for $q_0 = 0.5$): 
\begin{equation}
\Omega_{\rm HI}(z) = 0.19 \pm 0.04 \times 10^{-3} h^{-1} 
\exp(0.83 \pm 0.15 \times z) \,.
\label{eq:hyd}
\end{equation}
At higher redshift  a decrease in $\Omega_{\rm HI}$ has been reported (see for
instance Storrie-Lombardi et al. 1995) which suggests that the  formation of 
galaxies commenced around $z = 3$. For comparison with structure formation
models one needs to know the mass of damped Lyman-$\alpha$ systems
and this is highly uncertain. We follow Mo and Miralda-Escud\'e (1994)
(see also McNally \& Peacock 1996) in assuming that the 
minimum mass corresponding to these systems is $10^{9.6} h^{-1} M_\odot$
which corresponds to a virial  velocity of $\sim 50 \rm km \, sec^{-1}$. 
This limit comes from the smallest halos which could cool sufficiently
rapidly to collapse by $z \simeq 3$ (Efstathiou 1992). 
Knowing the minimum mass  one can use the  Press-Schechter formalism
(Press \& Schechter 1974)  to calculate the fraction of the total
matter  which could collapse in forming  structures with masses   $\ge
10^{9.6} h^{-1} M_\odot$ at any redshift $z$. This can be multiplied
with $\Omega_B$  to estimate the density of  baryons that has
collapsed into these objects and this gives  
\begin{equation}
\Omega_{\rm col}(M_{\rm min}, z) = \Omega_B \times \left (1 - {\rm erf}
\left [ {\delta_c (1+z) \over \sqrt(2)
\sigma(M_{\rm min}, 0)}\right ] \right)
\label{eq:coll}
\end{equation}

The only input from structure formation models  in  equation (\ref{eq:coll})
is the value of $\sigma(M_{\rm min}, 0)$  which is the r.m.s. mass
fluctuation  in  spheres of radius $R \, (\sim .3 \rm Mpc)$
corresponding to $M_{\rm min}$ evaluated  at the present epoch. The
uppermost curve in  figure  \ref{fig:sig} shows the value of $\sigma_R$
at a comparable  length-scale ( $R=.25 h \rm Mpc$) as a function of
the neutrino mass for a class of allowed models,  and this gives
an idea of how this quantity changes for different decaying neutrino
models.  We also use  $\delta_c = 1.686$ which is  the value
corresponding to the spherical collapse. 
It should be pointed out that equation (\ref{eq:coll}) gives the
collapsed fraction of baryons in all the structures with masses $\ge
M_{\rm min}$, though 
the masses of damped Lyman-$\alpha$ clouds lie between some $M_{\rm min}$
and $M_{\rm max}$, which probably correspond to the 
virial velocity $\sim 200 \, \rm km \, sec^{-1}$, comparable to the
present day galaxies.   The collapsed 
fraction is extremely insensitive to the  upper cut off in mass and  changes 
negligibly as $M_{\rm max}$ is changed from the mass corresponding to a 
virial  velocity of $\sim 200 \,  \rm km \, sec^{-1}$ to infinity which is the
value that  equation (\ref{eq:coll}) assumes. This is expected because
the probability of forming very high mass objects at high redshifts is 
exponentially suppressed. 

 An important point to note is that while $\Omega_{HI}$ in
equation (\ref{eq:hyd}) refers to the  neutral hydrogen observed in damped
Lyman-$\alpha$ clouds, the quantity being calculated here
i.e. $\Omega_{\rm col}$ in equation ({\ref{eq:coll}) refers to the total amount
of baryons that has collapsed into the damped Lyman-$\alpha$ clouds. 
It is expected that some of the  collapsed baryons will be in the form
of stars and ionized gas, and  therefore any acceptable model of
structure formation should predict  a value for $\Omega_{\rm col}$ which
is at least equal to  $\Omega_{\rm HI}(z)$ if not larger. 

In figure \ref{fig:dam} we show the collapsed fraction 
$\Omega_{\rm col}$ for some decaying  neutrino models. While the baryon
density  was ignored when calculating the matter power spectrum for
these models, we have used $\Omega_B = 0.05$ in equation
({\ref{eq:coll}). The shaded region in figure \ref{fig:dam} shows the
observed 
$\Omega_{\rm HI}$  (Eq.~(\ref{eq:hyd})) for $h = 0.5$ and the data point
at $z \simeq 4$ is taken from Storrie-Lombardi et al. 1995. For 
comparison we also show the collapsed fraction for a HCDM model with
$\Omega_\nu = 0.25$, where  $\Omega_\nu$ is  the fraction of the
matter in the hot component. The power spectrum for the HCDM model was
computed using COSMICS (Bertschinger \& Bode 1995) with the value
$\Omega_B = 0.05$.  As is clearly seen the HCDM model seems
to be at variance with observations as noted by several earlier
authors  (Mo \&  Miralda-Escud\'e 1994;  Padmanabhan  \& Subramanian
1994; Ma \& Bertschinger 1994;  McNally \& Peacock 1996; Ma et
al. 1997; Gardner et al. 1997). The two decaying neutrino models 
considered  here satisfy all the constraints discussed earlier in this
section and the power spectra for these models is  shown in  figure
\ref{fig:all}. The collapsed 
fraction for both  
the models exceeds the observed $\Omega_{\rm HI}(z)$ which indicates
that both these models are compatible with the damped Lyman-$\alpha$ cloud 
observations. Similar conclusions are expected to hold for the other
decaying neutrino models that pass the various tests discussed earlier
this section.  It should be pointed out that our computation of the
matter power spectrum  doesn't take into account the baryons. 
The inclusion of baryons reduces power at small scales (see e.g.  Hu \&
Sugiyama 1996) which means that the  computed fraction  shown in
figure \ref{fig:dam} is an overestimate by $15 \hbox{--} 30 \, \%$
depending on the value of $\Omega_B$. However, it is evident from
figure \ref{fig:dam} that  a decrement of this level in our computed
$\Omega_{\rm col}$  will not affect the conclusions of this
comparison with observations of damped Lyman-$\alpha$ clouds.. 

Similar constraints on structure formation models can also be derived from 
the recent observation of  high redshift galaxies (Steidel et al. 1996;
 Madau et al. 1996).  However, there is an even  greater uncertainty attached 
to the masses of these high redshift  galaxies and comparison with structure 
formation models may not be so  straight forward (Mo \& Fukugita 1996) as for
damped Lyman-$\alpha$ clouds. 

\section{Summary and Discussion}
We have studied large scale structure formation in  a $\Omega=1$
universe  with
decaying massive neutrinos. This variant of the CDM model has two
extra parameters---
the neutrino mass $m_{\nu}$ and the lifetime of the massive neutrino
$t_d$, and by varying these it is possible to introduce  
extra features into the power spectrum. We have computed the power
spectrum for a large range of the parameters $ m_{\nu}$ and $t_d$.
Our analysis takes into account the free-streaming of the massive
neutrinos, and this allows us to study a hitherto unexplored region of
parameter space corresponding to low neutrino masses and large
lifetimes.   

Unlike other models of  structure formation  in the universe - for
example the standard CDM model,  and its variants like $\rm HCDM$, $\lambda
\rm CDM$, $oCDM$ - the decaying neutrino model allows the
possibility  of introducing extra features in the power spectrum at
both  large scales ($k \le 0.01\, \rm Mpc^{-1}$) and small scales ($k
\ge 0.1 \, \rm Mpc^{-1}$). 
The decay of the neutrino acts to reduce the power at large scales (as
compared to CDM model) by increasing the radiation content of the
universe to a value which can be much larger than what is obtained
from  three  relativistic neutrino species. This reduces the power at
large scales by delaying the matter radiation equality. In addition,
the late radiation dominated era can affect the overall normalization
of the power spectrum through
the  `integrated Sachs-Wolfe effect' and we find that this can lead to
an overall reduction of power at all scales.  
The decaying neutrino models also allow us the possibility of enhancing the
power over a range of modes at  small scales,  and this is achieved by
varying the parameters to suitably adjust the first era of matter 
radiation equality. 

CDM-like models with $\Gamma$ between $0.22$ and $0.29$ seem to be
compatible  with most observations (Peacock \& Dodds 1994). To allow
for the uncertainties in comparing  the computed linear power
spectra with the  observed non-linear ones, i.e., from APM and LCRS
surveys, we allow a  larger range of $\Gamma$:  $0.2 \le \Gamma \le
0.3$. We find that it is possible to construct a large number of
models which 
predict a value of $\Gamma$ in this range and the allowed region of
parameter space is shown in figure \ref{fig:reg}.
We have studied in some detail  a class of models whose power spectrum
is similar to a $\Gamma=.24$ CDM power spectrum at large scales near
the turnaround from the primordial Harrison Zel'dovich form. We find
that for the mass range $m_{\nu} \ge 10  \,\rm keV$, for 
the entire  $k$ range  that we have studied  (i.e. $k \le 5 h \Mi$), the
power spectrum is indistinguishable from a CDM power spectrum with
$\Gamma \simeq 0.24$. As the mass is reduced below $10 \, \rm keV$ the
power spectrum starts getting extra power at small scales and the
shape of the power spectrum starts to differ considerably from a
CDM-like power spectrum with a $\Gamma$ fit.   We  find that the
models in the mass range $m_{\nu} > 50 \rm eV$ are roughly consistent
with the APM and LCRS power spectrum, and  with $\sigma_8$ inferred
from cluster abundances, but with extra power at small scales. 
We find that the predicted r.m.s. peculiar velocities in
spheres of radius  $40 h^{-1} \rm Mpc$ and $60 h^{-1} \rm Mpc$ 
are somewhat below those indicated by observations, but the
discrepancy is not as severe so as to conclusively rule out these
models.  For masses  $ \le 50 \, \rm eV$ the values predicted
for $\sigma_8$ are too large to be compatible with observations of
cluster abundances. This conclusion is independent of the choice of
$t_d$ and we fail to find any value of $t_d$ for which models with   
 $m_{\nu} \le 50 \rm eV$ produces an acceptable power spectrum. 

We also find that for  very small masses  the
neutrinos have to decay very late if this process is to have a
significant effect  on  the power spectrum. In these models the power
spectrum has too much power at small scales and too little power at
large scales, and these models are ruled out by observations
(figures \ref{fig:all}, \ref{fig:nal}   and \ref{fig:dec}, and Table~1). 

We have compared the predictions of some of the allowed decaying
neutrino models with the observations of the abundances of damped
Lyman-$\alpha$ clouds at high redshifts and  we find that the decaying
neutrino models are compatible with these observations.

We have addressed the question if any of the radiatively decaying
neutrino  models which can reionize the intergalactic medium is
consistent with the observed large scale structure of the universe. We
fail to find any model which can reionize the IGM and also
produce an acceptable power spectrum in a $\Omega=1$ universe. It is
still to be seen if such models can be constructed by considering open
universes or by introducing a cosmological constant. In addition to
the constraints on radiatively decaying neutrinos considered in this
paper, models which predict very early reionization are also
constrained by CMBR observations at small angular scales (Netterfield
et al. 1997). However, as we find that these models are ruled
out by the structure formation considerations presented in this paper,
we have not considered the small angle CMBR observations here. 
Incidentally, an
early epoch of enhanced galaxy formation may be expected in the
allowed models with extra small scale power, and this may provide a
method of reionizing the IGM. Issues related to galaxy formation
in a  decaying neutrino model are beyond the scope of this paper, and
they require a more detailed analysis involving N-body simulations and
baryonic physics. Such an analysis will also be able to put further
restrictions on the allowed regions of the $m_{\nu}$--$t_d$ parameter
space.  In addition, N-body simulations for the decaying neutrino
power spectra will permit comparison with the observed power spectrum
at scales where the non-linear effects are important and this can be
used to put  restrictions on the allowed models. 

        Finally, although we see that it is possible to use the
observed distribution of matter in the universe to restrict the
parameter space of decaying neutrino models,  it seems 
implausible that these observation will be able to distinguish
a decaying neutrino model from other variants of the CDM model in the
near future. However, the study of CMBR anisotropies at small
angular scales holds the promise of being able to discriminate between
these models and possibly single out the correct one. 
 Two satellite projects planned for the next decade ---MAP and Planck
surveyor (COBRAS-SAMBA)---will  map the CMBR sky at a few arc-minutes
angular resolutions, and it is hoped that these observations will
unambiguously validate one of the models and reject the rest. In light
of this it important to understand the small angular scale
anisotropies in the CMBR for the various allowed decaying neutrino
models, and work is in progress in this direction. 

\acknowledgements
The authors would like to thank Tarun Souradeep for useful discussions.
S.B. would like to thank Rajaram Nityananda for some useful
discussions, and the Raman Research Institute for the use of its
computer facility in the early stages of this project. S.K.S.  would
like to thank Profs. J. R. Bond and J. Silk for useful comments and
suggestions. The authors thank the referee Martin White for useful
comments and suggestions. 

\appendix
\section{Formalism}
Here we briefly discuss the equations that govern the
evolution of the scale factor and linear perturbations in a spatially flat
universe which 
is composed of CDM particles, photons tightly coupled to baryons, massless
neutrinos and  massive neutrinos which have the possibility of
decaying into some massless (relativistic) decay products.  The
evolution of both the background universe and the perturbations
is governed by the Einstein equations 
\be
R^{\mu}_{\nu}=8 \pi G (T^{\mu}_{\nu}-
\frac{1}{2} \delta^{\mu}_{\nu}T )  \,. \label{eq:a1}
\ee
where the Ricci tensor $R^{\mu}_{\nu}$ is calculated from the metric,
and the energy momentum tensor $T^{\mu}_{\nu}$ has contributions from
all the different particle species. In this part of the paper we use $c=1$. 

In the synchronous gauge the metric components can be written as  
\be 
g_{00}=-a^2(\e) \hspace{.5cm} g_{oi}=0 \hspace{.5cm}
g_{ij}=  a^2(\e) \lsb \delta_{ij}+h_{ij}(\x,\e) \rsb \label{eq:a2}
\ee
where the zeroth component of the coordinates refers to the conformal
time $\e$ and $h_{ij}$ is the metric perturbation. Using this
in the Einstein equation one obtains the equation for the scale factor
and the metric perturbation. Using prime to denote  derivative with
respect to $\e$,  the equation for the scale factor  can be written as
\be
\ap(\e)=H_0 \sqrt{\om(\e)}\;. \label{eq:a3} 
\ee

This equation is discussed in some detail in section 2. 
We next consider the equations for the evolution of $h^i_j$ the metric
perturbation.   We proceed by first calculating  the perturbation in
the Ricci tensor caused by the metric perturbation.
 We only use the components $R^0_0$  and $R^0_j$, and keeping only
terms linear in $h^i_j$, we obtain  
\be
\D R^0_0= \frac{1}{2 a^2}\lb h^{''}+\frac{a^{'}}{a} h^{'} \rb 
\label{eq:a4}
\ee
and
\be
\D R^0_j=  \frac{1}{2 a^2}\lb h^{'}_{,j}-
h^{'l}_{j,l} \rb  \,. \label{eq:a5}
\ee
Here $h$ is the trace of the metric perturbation, and we have used the
notation  ${}_{,j} \equiv \frac{\pl}{\pl x^j}$.

        It is possible to decompose the metric perturbation into
scalar, vector and tensor components, and at the linear order these evolve
independently. Retaining only the scalar parts, we  write the metric
perturbation as  
\be h^i_j(\x,\e)=\int \frac{d^3 k}{(2 \pi)^3} e^{i \k \cdot \x}
\frac{1}{3}  \lsb \mu(\k,\e) \delta^i_j + \la(\k,\e) \lb
\delta^i_j - 3 \frac{k^i k_j} {k^2} \rb  \rsb \label{eq:a6}
\ee 
where $\mu$ which is the Fourier transform of $h$ corresponds to
isotropic dilations or contractions, and $\la$, which is the traceless
part, corresponds to shear. Expressing 
the Fourier transform of the curvature in terms of these quantities we have 
\be
 \D {\tilde R}^0_0(\k,\e)=  \frac{1}{2 a^2}\lb \mu^{''}+\frac{a^{'}}{a}
\mu^{'} \rb \label{eq:a9} 
\ee
and 
\be
 \D {\tilde R}^0_j(\k,\e)=  \frac{i k_j}{3 a^2}\lb \mu^{'}+
\la^{'} \rb \label{eq:a10} 
\ee

We also write  the Fourier transform of the linear order perturbation
to the energy momentum tensor as
\be
\D {\tilde T}^0_0(\k,\e)= - \D \rho(\k,\e) \hspace{1.cm}
\D {\tilde T}^i_j(\k,\e)= \D P(\k,\e) \delta^i_j + \frac{1}{2} \D
Q(\k,\e) (\delta^i_j-3  \frac{k^i k_j}{k^2}) \,. \label{eq:a11}
\ee 
where $\D \rho(\k,\e),\D P(\k,\e) {\rm and} \D Q(\k,\e)$ are  the
Fourier transform of the perturbations in the density, the pressure
and the anisotropic stresses respectively.  The perturbations to the 
components $T^0_j$ also have contributions at linear order, but as
discussed below, the final equations that we use do not have any
explicit reference to $\D {\tilde T}^0_j$.    

        Using the above expressions in the Einstein equation, the ${}^0_0$
component gives us an equation for $\mu$
\be
\mu^{''}+\frac{a^{'}}{a} \mu^{'} = - 3 H_0^2 a^2 \lb \frac{\D \rho}{\rc} +3
\frac {\D P}{\rc} \rb \,. \label{eq:a15}
\ee

To follow the evolution of $\la$ we use the ${}^0_j$ component of the
Einstein equation which gives us
\be
\mu^{'}+\la^{'}=  \frac{-9 H_0^2 a^2 }{\rho_{c0}} \frac{i k^j}{k^2}
 \D {\tilde T}^0_j \,. 
\ee
Differentiating  this once with respect to $\e$ and using the
energy-momentum conservation equation $T^{\mu}_{j;\mu}=0$,  we obtain
the following equation for $\la$
\be 
\la^{''}+\frac{a^{'}}{a} \lb 2 \la^{'} + \mu^{'} \rb =
3 H_0^2 a^2 \lb \frac{\D \rho}{\rc}  + 3 \frac{ \D Q }{\rc} \rb
\,. \label{eq:a16} 
\ee 

Finally, we have the  equations (\ref{eq:a3}),(\ref{eq:a15}) and
(\ref{eq:a16}) which we use to follow the evolution of $a$,$\mu$ and
$\la$ respectively. The right hand side of these equations has
quantities which refer to the total density, pressure and anisotropic
stresses. These have contributions from the different
species of particles and we have to consider 
them separately. The total effect is obtained by summing over the
contribution from the different species e.g. $\om=\sum
\om_{species}$,$\D \rho =\sum \D \rho_{species}$,  etc. In the
following subsections we separately consider the different
kinds of species we have taken into account. 

\subsection{The ideal fluids.}

Cold dark matter  particles and photons which are tightly coupled to
baryons can be treated as ideal fluids. The energy momentum tensor
then has the form 
\be
T^{\mu}_{\nu}=P \delta^{\mu}_{\nu} + (P +\rho) U^{\mu} U_{\nu}
\ee
where $U^{\mu}(x,\e)$ is the bulk 4 velocity of the fluid, and there
are no anisotropic stresses i.e. $Q=0$.  

The CDM particles can be considered as  dust for which there is no
pressure i.e. $P_{\cdm}=0$.  We choose a synchronous 
coordinate system which moves with the dust 
particles and hence these particles have no peculiar velocities
i.e. $U_{\cdm}^{\mu}=(1/a,0,0,0)$. The energy-momentum conservation equation 
${T_{\cdm}\hspace{.1cm}}^{\mu}_{0;\mu}$ for  the CDM component then gives
the equation   
\be
\om_{\cdm}(\e)= a(\e) \Om_{\cdm 0}
\ee
for the background density, and the equation 
\be
\d^{'}_{\cdm}(\k,\e)+\frac{1}{2} \mu^{'}(\k,\e) = 0 \label{eq:x1}
\ee
 with 
$$
\frac{\D \rho_{\cdm}}{\rc}=\frac{\om_{\cdm} \d_{\cdm}}{a^4} \,, 
$$
for the density perturbation.

The photon-baryon fluid has the equation of state $P_{\g}=(1/3)
\rho_{\g}$.  
For this fluid the energy momentum conservation equations give the
equation 
\be
\om_{\g}(\e)=  \Om_{\g\hspace{.1cm}0}
\ee
for the background density. Perturbations in this medium produce bulk
flows relative to the synchronous coordinate system,
and in addition to  the perturbation in the density, one has to also
consider the  perturbation to the  velocity of the fluid $\D
U^{i}_{\g}(x,\e)$. Only the divergence of this quantity
$(\theta_{\g}= \D U^{i}_{\g,i})$ couples to the density
perturbation, and the energy-momentum conservation gives us the
equations    
\be
\d^{'}_{\g}(\k,\e)+\frac{4}{3} \tilde{\theta_{\g}}(\k,\e)+ 
\frac{2}{3} \mu^{'}(\k,\e)=0 \label{eq:x2}
\ee
and
\be
\tilde{\theta_{\g}}^{'}(\k,\e)- \frac{k^2}{4} \delta_{\g}(\k,\e)=0
\label{eq:x3}
\ee
which we use to follow the evolution of $\D \rho_{\g}/\rc=\om_{\g}
\d_{\g}/a^4$ and $\D P_{\g}/\rc=\om_{\g} \d_{\g} /(3 a^4)$.

\subsection{Neutrinos.}
It is not possible to treat neutrinos as a perfect fluid and a
microscopic description is required. Every particle is fully described
by its position in space-time and its momentum. 
Instead of using the momentum components in the synchronous coordinate
system, it is more convenient to use the components of the momentum
on the tetrad
\be  
\eb_{b}=\frac{1}{a^2}(\d_b^{\mu}-\frac{1}{2} h_{b}^{\mu})
\frac{\pl}{\pl x^{\mu}} \hspace{.5cm}
\eb^{b}=a^2(\d_{\mu}^b+\frac{1}{2} h^b_{\mu}) d x^{\mu} \label{eq:c1}
\ee
introduced by Bond \& Szalay (1983). Here the tetrad index $b$ takes
values $0,1,2,3$, and the tetrad is orthogonal (but not normal).
The metric has  components $g_{bc}={\bf g}(\eb_b.\eb_c)=a^{-2}
\eta_{bc}$ on the 
tetrad, and a particle's  4-momentum ${\bf p}$ has components  ${\bf
p}=q^a \eb_a$ on the tetrad. The components $q^a$ are 
related by $q^a q^b \eta_{a b}=-a^2 m^2$, and any 3 of the 4 components
are sufficient to fully describe the momentum state of a particle.
We use the 3 spatial components $q^i$ and the zeroth
component is obtained from these using $q^0=\sqrt{q^2+a^2 m^2}$, where
$q^2$ is used to denote $\d_{ij} q^i q^j $. 

 Next,  the  equation of motion (parallel transport) ${\bf
\nabla}_{\bf 
p} {\bf p}=0$ for a particle  is used to  arrive at  the equation for
the evolution of $q^a$
\be
q^b \eb_b(q^a)=- \lsb({\bf \nabla}_{\eb_b}\eb_c)(\eb^a) \rsb q^b q^c\,.
\label{eq:c2}
\ee
Keeping only terms linear in the metric perturbation equation
(\ref{eq:c2}) gives  
\be
q^b \eb_b(q^i)=-\frac{1}{2 a^2} q^b q^c (h^i_{b,c}-{h_{bc,}}^i) \,  
\label{eq:c3}
\ee
for the spatial components of the momentum  $q^i$, 
and it follows that in the absence of any perturbations the 3 spatial 
components of the momentum remain constant, and they
evolve only due to the perturbation in the metric.

 The state of a large number of neutrinos of a particular 
species can be described by a distribution function $f(\x,\q,\e)$
which is the number density of these particles in phase space, and its
evolution equation  
\be
q^a \eb_a(f) + q^a \eb_a(q^i) \frac{\pl}{\pl q^i} f = 0 \label{eq:c4}
\ee 
follows from the local conservation of particles in phase space. 
We also consider situations where the particles of a particular species
decay to produce particles of a different species, and for such
situations it is necessary to introduce a source term  in equation
(\ref{eq:c4}). This is  discussed later in this section. 

        The distribution function is next decomposed into two parts,
an isotropic function of $q$  corresponding to the distribution of
particles in the unperturbed universe,  and another corresponding to
the perturbation i.e.  
\be 
f(\x,\q,\e)=\fb(q,\e)+ \d f(\x,\q,\e) \,.
\ee
Equation (\ref{eq:c4}) then gives 
\be
\frac{\pl}{\pl \e} \fb(q,\e)=0 \hspace{.3cm} {\rm i.e.}  \hspace{.3cm}
\fb(q,\e)=\fb(q) \label{eq:c5}
\ee
for the unperturbed distribution function. For neutrinos the
unperturbed distribution function is the  Fermi-Dirac distribution
function
\be
\fb_{\nu}(q)=\frac{2}{h_P^3 \lsb \exp(q/k_B T_{\nu 0})+1 \rsb}
\label{eq:cc5} 
\ee
where the factor 2 takes into account the fact that for every neutrino
species there will be both particles and anti-particles, $h_P$ is the
Planck constant, $k_B$ is the Boltzmann constant and $T_{\nu 0}=2.726
^o K /1.4$ is the present temperature of the relic cosmic  neutrinos.

For the perturbation, equation (\ref{eq:c4}) gives us 
\be 
\frac{\pl}{\pl \e} \d f + \frac{q^i}{q^0} {\d f}_{,i}-\frac{q^i q^j}{2
q}  h_{ij}^{'} \dq \fb =0 \,. \label{eq:c6}
\ee 
Using $F(\k,\q,\e)$ to denote the Fourier transform of
$\d f(\x,\q,\e)$ and using $\a$  to denote the cosine of the angle between
$\q$ and $\k$, equation (\ref{eq:c6}) can be written as  
\be 
\frac{\pl}{\pl \e} F + \frac{i \a q k}{q^0} F - \frac{q}{6} \lsb
\mu^{'} +(1-3 \a^2) \la^{'} \rsb  \dq \fb =0 \,. \label{eq:c7}
\ee 
and we use these equations to follow the evolution of  the energy
momentum tensor 
\be
T^a_b=\frac{\e_{b c}}{a^4} \int \frac{q^a q^c}{q^0} f d^3 q
\,.\label{eq:c8} 
\ee
The  background density can be written in terms of the distribution
function as  
\be
\rho(\e)= \frac{1}{a^4} \int  q^0 {\bar f}(q,\e)  d^3 q \,, \label{eq:c9}
\ee
and the various perturbed quantities that appear in the equation for
$\mu$ and $\la$ can be written as 
\be
\D \rho(\k,\e)=\frac{1}{a^4} \int  q^0  F(\k,\q,\e) d^3 q \,,\label{eq:c10}
\ee
\be
\D P(\k,\e)=\frac{1}{3 a^4} \int  \frac{q^2}{q^0} F(\k,\q,\e) d^3 q
\label{eq:c11}
\ee
and
\be
\D Q(\k,\e)=\frac{1}{3 a^4} \int (1-3 \alpha^2) \frac{q^2}{q^0}
F(\k,\q,\e) d^3 q  \,. \label{eq:c12} 
\ee
\subsubsection{Massless neutrinos.}
For massless neutrinos the calculation is greatly simplified because
$q^0(q,\e)=q$ which does not evolve in time. The integral in equation
(\ref{eq:c9}) is a constant and 
as a result  $\om_{\nu}(\e)=\Om_{\nu 0}$.  Equation (\ref{eq:c7}) is
solved by the method of characteristics and the solution is
\bea 
F_{\nu}(\k,\q,\e)&=&F_{\nu}(\k,\q,\e_i)e^{-i \a k (\e-\e_i)}  \\
&+& \frac{q}{6} \dq \fb_{\nu} \ino \lsb \mp(\k,\ep)+(1-3
\a^2)\lp(\k,\ep) \rsb e^{-i \a k  (\e-\ep)} d \ep \nonumber\,
\label{eq:c13} 
\eea
where $\e_i$ is the instant when the initial conditions are specified
and from which we start following the evolution of the perturbations.  
We use this  in equations (\ref{eq:c10}), (\ref{eq:c11}) and
(\ref{eq:c12}) where the $d^3 q$ integral can be done analytically. The
angular integrals involve the following relation involving Legendre
polynomials $P_l(x)$ and spherical Bessel functions $j_l(x)$
\be
j_l(x)=\frac{i^l}{2} \int^1_{-1} e^{i \a x} P_l (\a) d \a \,,
\label{eq:c14} 
\ee
and for $l \ge 2$ we also define 
\bea
\F_l(x)&=& \frac{1}{ 8 l^3 + 12 l^2 - 2 l -3} 
\lsb (6 l^3 + 15 l^2 + 3 l -6) j_{l+2}(x)- (4 l^3 + 6 l^2 + 2 l)
j_l(x)  \nonumber \right. \\ 
&+& \left. (6 l^3 + 3 l^2 - 9 l) j_{l-2}(x) \rsb
= \frac{i^l}{2} \int^1_{-1} e^{i \a x} (1-3 \a^2)^2 P_l(\a) d \a
 \label{eq:c15} \,.
\eea  
We use these to obtain
\bea
\frac{\D \rho_{\nu}(\k,\e)}{\rc}&=&3 \frac{\D P_{\nu}(\k,\e)}{\rc} = 
-\frac{2 \om_{\nu}}{3 a^4} \left\{ \lsb \mu(\k,\e_i) j_0(k(\e-\e_i)) +
2 \la(\k,\e_i) j_2(k(\e-\e_i)) \rsb \right. \nonumber \\ 
&+& \left.  \ino \lsb \mp(\k,\ep) j_0(k(\e-\ep)) + 2 \lp(\k,\ep)
j_2(k(\e-\ep)) \rsb d \ep \right\} \label{eq:c16} 
\eea
and 
\bea
\frac{\D Q_{\nu}(\k,\e)}{\rc}&=& -
\frac{4 \om_{\nu}}{3 a^4} \left\{ \lsb  \mu(\k,\e_i) j_2(k(\e-\e_i)) + 
\la(\k,\e_i) \F_2(k(\e-\e_i)) \rsb \right. \nonumber \\ 
&+& \left.  \ino \lsb  \mp(\k,\ep) j_2(k(\e-\ep)) + \lp(\k,\ep)
\F_2(k(\e-\ep)) \rsb d \ep \right\}  \label{eq:c17} \,.
\eea
It should be noted that in obtaining equations (\ref{eq:c16}) and 
(\ref{eq:c17})  we have used a particular form for the initial perturbation
$F(\k, \q,\e_i)$ which has been chosen such that it  corresponds to
the growing mode of the perturbation, and this is discussed in more
detail later.  

We use equation (\ref{eq:c16}) and (\ref{eq:c17}) to follow the
evolution of perturbations in the massless neutrinos.  

\subsubsection{Decaying massive neutrinos.}
For decaying massive neutrinos a source term has to be included in
equation (\ref{eq:c4}) and we have 
\be
q^a \eb_a(f_d) + q^a \eb_a(q^i) \frac{\pl}{\pl q^i} f_d = - \frac{\m}{t_d}
f_d  
\label{eq:d1}
\ee 
where $t_d$ is the lifetime of the massive neutrinos. The evolution
of the unperturbed distribution function is governed by 
\be
\frac{q^0}{a^2} \frac{\pl}{\pl \e} \fb_d(q,\e)=- \frac{\m}{t_d} \fb_d
(q,\e) \,. \label{eq:d2} 
\ee
We write the solution of this equation in terms of the variable
\be
\psi(q,\e)= \int^\e_{\e_i} \frac{ \m a^2 (\ep) }{q^0(q,\ep)} d \ep
\,. \label{eq:d3} 
\ee
which corresponds to the proper time of a neutrino with spatial
momentum  $q$, and it goes over to the cosmological time $t$ in the
limit $q \ll a \m $. This variable takes into account the fact that the
decay of the neutrino is governed by the passage of time in its own
rest frame,  and not in the frame of the cosmological observer. The
solution for the unperturbed distribution function is 
\be
\fb_d(q,\e)= e^{- \psi(q,\e)/t_d} \fb_{\nu}(q)\,. \label{eq:d4}
\ee
where $\fb_{\nu}(q)$ is the Fermi-Dirac distribution given in
equation (\ref{eq:cc5}). We use equation (\ref{eq:d4}) to follow the
evolution of the background density of the massive neutrinos and we
have 
\be 
\rho_d(\e)=\frac{1}{a^4} \int d^3 q q^0(q,\e) \fb_d(q,\e)\,. \label{eq:d5a}
\ee
The  equation for the perturbation is 
\be 
\frac{\pl}{\pl \e} \d f_d + \frac{q^i}{q^0} {\d f_d}_{,i}-\frac{q^i q^j}{2
q}  h_{ij}^{'} \dq \fb_d =- \frac{\m a^2}{t_d q^0} \d f_d  \label{eq:dd5}
\ee 
which can be simplified by defining 
\be 
\d f_d(\x,\q,\e)= e^{-\psi/t_d} \d {\hat f}_d(\x,\q,\e) \label{eq:dd6}
\ee
and 
\be
\dq \bg_d(q,\e)  = \dq \fb_{\nu}(q) - \lsb \frac{1}{t_d} \dq 
\psi(q,\e)  \rsb  \fb_{\nu}(q) \label{eq:dd7} \,,
\ee 
and equation (\ref{eq:dd5}) can then be written as 
\be 
\frac{\pl}{\pl \e} \d {\hat f}_d + \frac{q^i}{q^0} {\d {\hat
f}_d}_{,i}- \frac{q^i q^j}{2 q}  h_{ij}^{'} \dq \bg_d =0 \,. \label{eq:d5}
\ee 

Using $\Fh_d(\k,\q,\e)$ to denote the Fourier transform of $\d {\hat
f}_d(\x,\q,\e)$, the solution of equation (\ref{eq:d5}) can be written
in Fourier space as 
\bea
\Fh_d(\k,\q,\e) &=&\Fh_d(\k,\q,\e_i)e^{-i \a k  \t}  \nonumber \\
&+& \frac{q}{6} \ino  \dq \bg_d(q,\ep) \lsb \mp(\k,\ep)+(1-3
\a^2)\lp(\k,\ep) \rsb 
e^{-i \a k  (\t-\tp)} d \ep  \label{eq:d6} 
\eea
where the variable
\be 
\t(q,\e)=\ino \frac{q}{q^0(q,\e_1)} d \e_1 \label{eq:d7}
\ee
corresponds to the comoving distance a neutrino travels in the time
interval $(\e-\ep)$, and we use $\tp$ to denote $\tau(q,\ep)$. For
massive neutrinos it is not possible to do the $q$  integral
analytically. Doing the angular integrals analytically we  obtain 
\bea
\D \rho_{d}(\k,\e) &=&
\frac{2 \pi}{3 a^4}\int d q q^3 q^0 e^{-\psi/t_d}
 \left\{ \dq \fb_{\nu}(q) \lsb \mu(\k,\e_i)
j_0(k \t) + 2 \la(\k,\e_i) j_2(k \t) \rsb \right. \nonumber \\ 
&+& \left.  \ino \dq \bg_d(q,\ep)  \lsb \mp(\k,\ep) j_0(k(\t-\tp)) +
2\lp(\k,\ep) j_2(k(\t-\tp)) \rsb d \ep \right\} \,, \nonumber \\
\label{eq:d8} 
\eea
\bea
3 \D P_{d}(\k,\e) &=&
\frac{2 \pi}{3 a^4}\int d q \frac{q^5}{q^0} e^{-\psi/t_d}
 \left\{ \dq \fb_{\nu}(q) \lsb \mu(\k,\e_i)
j_0(k \t) + 2 \la(\k,\e_i) j_2(k \t) \rsb \right. \nonumber \\ 
&+& \left.  \ino \dq \bg_d(q,\ep)  \lsb \mp(\k,\ep) j_0(k(\t-\tp)) +
2\lp(\k,\ep) j_2(k(\t-\tp)) \rsb d \ep \right\} \,, \nonumber \\
\label{eq:d9} 
\eea
and
\bea
3 \D Q_d(\k,\e) &=&
\frac{4 \pi}{3 a^4}\int d q \frac{q^5}{q^0} e^{-\psi/t_d}
 \left\{ \dq \fb_{\nu}(q) \lsb \mu(\k,\e_i)
j_2(k \t) +  \la(\k,\e_i) \F_2(k \t) \rsb \right. \nonumber \\ 
&+& \left.  \ino \dq \bg_d(q,\ep)  \lsb \mp(\k,\ep) j_2(k(\t-\tp)) +
\lp(\k,\ep) \F_2(k(\t-\tp)) \rsb d \ep \right\} \,. \nonumber \\
\label{eq:d10} 
\eea
We also use equation (\ref{eq:d6}) to follow the evolution of 
\be
\D n_d(\k,\e)=\int F_d(\k,\q,\e) d^3 q
\ee
and we obtain 
\bea
\D n_d(\k,\e) &=&
\frac{2 \pi}{3}\int d q q^3 e^{-\psi/t_d}
 \left\{ \dq {\hat f}_d(q) \lsb \mu(\k,\e_i)
j_0(k \t) + 2 \la(\k,\e_i) j_2(k \t) \rsb \right. \nonumber \\ 
&+& \left.  \ino \dq \bg_d(q,\ep)  \lsb \mp(\k,\ep) j_0(k(\t-\tp)) +
2\lp(\k,\ep) j_2(k(\t-\tp)) \rsb d \ep \right\} \,, \nonumber \\
\label{eq:d11} 
\eea
for this quantity which is relevant later for the study of
perturbations in the  decay product. 

We use equations (\ref{eq:d8}), (\ref{eq:d9}) and (\ref{eq:d10}) to
follow the evolution of perturbation in the massive decaying
neutrinos. 

\subsubsection{Decay product.}
Consider the decay of a massive neutrino which has  momentum $\q$ and
hence is in motion relative to the observers who define the
synchronous coordinate system.  In its own rest frame the  neutrino
decays by emitting 2 massless particles in exactly opposite directions
$\l$ and $-\l$,  and the direction $\l$ is  isotropically
distributed. We use $\Q_1(\q,\l)$ to denote the tetrad components of the
momentum of the decay particle which was emitted in the direction $\l$
in the neutrino's rest frame. Although the $\l$'s have an isotropic
distribution, the $\Q_1(\q,\l)$s will not be isotropically distributed,
and this happens because of the transformation from the the neutrino's
rest frame to the tetrad. To avoid confusion with the phase space of
the massive neutrino, we use coordinates $(\x,\Q,\e)$ on the phase
space of the decay product particles, and we add a source term to
equation (\ref{eq:c4}) to follow the evolution of the decay products. 
This extra term takes into account the fact that massive neutrinos
with all possible momenta $\q$ decay in all possible directions $\l$
(in the respective rest frame)
to give rise to 2 decay product particles, and the equation for the
evolution of the decay product is 
\bea
Q^a \eb_a(f_R(\x,\Q,\e)) &+& Q^a \eb_a(Q^i) \frac{\pl}{\pl Q^i}
f_R(\x,\Q,\e) = \nonumber \\& &
\frac{\m}{t_d} \int d^3 q  \frac{d \Om_{\l}}{4 \pi} 
\d^3(\Q-\Q_1(\q,\l)) f_d(\x,\q,\e) \label{eq:p1} \,.
\eea 

For the unperturbed distribution function this gives us 
\be
Q \frac{\pl}{\pl \e}  \fb_R(Q,\e)=\frac{a^2 \m}{t_d} \int d^3 q \frac{d
\Om_{\l} }{4 \pi} \d^3(\Q-\Q_1(\q,\l)) \fb_d(q,\e) \label{eq:p2} \,.
\ee
which on integrating over $d^3 Q$ gives 
\be
\frac{\pl}{\pl \e} \om_R(\e)=\frac{a^2 \m }{\rc t_d} \int d^3 q \fb_d(q,\e)
\label{eq:p3} 
\ee
for the background density of the decay product. For the perturbation
we have 
\bea
\frac{\pl}{\pl \e} \d f_R(\x,\Q,\e)&+&\frac{Q^i}{Q} {\d f_R}_{,i}(\x,\Q,\e) -
\frac{Q^i Q^j}{2 Q} h^{'}_{ij} \frac{\pl}{\pl Q} \fb_R(Q,\e)= \nonumber \\& &
\frac{a^2 \m}{Q t_d} \int d^3 q  \frac{d \Om_{\l}}{4 \pi} 
\d^3(\Q-\Q_1(\q,\l)) \d f_d(\x,\q,\e)  \,.
\label{eq:p4}
\eea 
The initial condition for the decay product is different (i.e. $\d
f_R(\x,\Q,\e) =0$) as we assume
that the initial density of the decay products is zero.  Using this,
the solution to equation (\ref{eq:p4}) can be written in Fourier
space as 
\bea 
F_R(\k,\Q,\e)&=& \frac{Q}{6}  \ino \dQ \fb_R(Q,\ep) \lsb \mp(\k,\ep)+(1-3
\a^2)\lp(\k,\ep) \rsb  e^{-i \a k (\e-\ep)} d \ep  \nonumber \\ &+&
\frac{\m}{Q t_d} \int d^3 q  \frac{d \Om_{\l}}{4 \pi}  
\d^3 (\Q-\Q_1(\q,\l)) \ino a^2(\ep) F_d(\k,\q,\ep) 
e^{-i \a k  (\e-\ep)} d \ep  \label{eq:p5}
 \nonumber \\
\eea
where now $\a$ is the cosine of the angle between $\Q$ and $\k$, and
$F_d(\k,\q,\e)= e^{-\psi/t_d} {\hat F}_d(\k,\q,\e)$. We
use this to calculate the density perturbation for which we obtain

\bea
\frac{\D \rho_{R}(\k,\e)}{\rc}&=&3 \frac{\D P_{R}(\k,\e)}{\rc} = 
-\frac{2 \om_{R}}{3 a^4} \ino \lsb \mp(\k,\ep) j_0(k(\e-\ep)) + 2
\lp(\k,\ep) j_2(k(\e-\ep)) \rsb d \ep 
 \nonumber \\
&+&\frac{\m}{\rc a^4 t_d} \int d^3 q  \frac{d \Om_{\l}}{4 \pi}  
 \ino a^2(\ep) F_d(\k,\q,\ep) 
e^{-i \a_1(\q,\l) k  (\e-\ep)} d \ep  \label{eq:p6a}
\eea
where $\alpha_1(\q,\l)$ is the cosine of the angle between $\Q_1(
\q,\l)$ and $\k$.  Doing the integral keeping the $\q$ and $\l$
dependence of $\a_1(\q,\l)$ is rather complicated. 
The calculation is greatly simplified if we assume that the neutrino
is at rest when it decays. Under this assumption the angle $\a_1$ is
the cosine of the angle between $\l$ and $\k$ and it no longer depends
on $\q$. This assumption is  reasonably good in situations where the
neutrino decays much after it has become nonrelativistic.

Under this assumption it is possible to  analytically do the angular
integral $d \Om_{\l}$,  and we obtain   
\bea 
\frac{\D \rho_R(\k,\e)}{\rc}&=&3 \frac{\D P_R(\k,\e)}{\rc} = 
-\frac{2 \om_{R}}{3 a^4} \ino \lsb \mp(\k,\ep) j_0(k(\e-\ep)) + 2
\lp(\k,\ep) j_2(k(\e-\ep)) \rsb d \ep 
 \nonumber \\
&+& \frac{\m}{\rc a^4 t_d}   \ino a^2(\ep) j_0(k(\e -\ep))
\d n_d(\k,\ep) d \ep  \label{eq:p6}
\eea
and 
\bea 
\frac{\D Q_R(\k,\e)}{\rc}&=& -\frac{4 \om_{\nu}}{3 a^4} 
\ino \lsb  \mp(\k,\ep) j_2(k(\e-\ep)) + \lp(\k,\ep)
\F_2(k(\e-\ep)) \rsb d \ep   \nonumber \\
&+& \frac{2 \m}{\rc a^4 t_d}   \ino a^2(\ep)  j_2 (k(\e -\ep))
\d n_d(\k,\ep) d \ep \,. \label{eq:p7} 
\eea
where
\bea
\d n_d(\k,\e)= \int F_d(\k,\q,\e) d^3 q \,.  \nonumber
\eea
The evolution of $\d n_d(\k,\ep)$ is governed by equation
(\ref{eq:d11}) which has been obtained in the previous
subsection. . We use equations (\ref{eq:p6}) and (\ref{eq:p7}) to 
follow the evolution of perturbations in the relativistic decay
product. 
\section{Initial conditions.}
Here we briefly discuss the initial conditions for adiabatic
perturbations. 
The initial conditions are set at an early epoch when the universe is
dominated by the  
relativistic  particles i.e. photons and neutrinos, and the massive
neutrino behaves like a relativistic particle and can also be treated
as a massless species. 
The scale factor then evolves as 
\be 
a(\e) = \e H_0 \sqrt{\Om_{\g0} +(n_{\nu}+1)\Om_{\nu0}} \label{eq:in1}
\ee
where $n_{\nu}$ is the number of massless neutrino species, and
$\Om_{\nu 0}$ is the contribution that one  massless neutrino species
would make to the present value of $\Om_0$. We also introduce the
quantity 
\be
r_{\nu}=\frac{(n_{\nu}+1)\Om_{0 \nu}} {\Om_{0 \g} +(n_{\nu}+1)\Om_{0 \nu}} 
\ee
which is the ratio of the density of the neutrinos to the total
density of the universe in the radiation dominated era. 

The initial epoch is also chosen such that all the relevant modes
are outside the horizon i.e. $k \e_i \ll 1$. In this limit the
equations for the evolution of perturbations in the photons and
the neutrinos are quite  simple and can be analytically solved. 
In the limit  $k \e_i \ll 1$ the equation for the perturbation in the
photon-baryon fluid  becomes 
\be
\d^{'}_{\g}(\k,\e))+ \frac{2}{3} \mu^{'}(\k,\e)=0 \,. \label{eq:i2}
\ee
For each neutrino species we define $\D \rho_{\nu}(\k,\e)=3 \D
P_{\nu}(\k,\e) = \rho_{\nu}(\e) \delta_{\nu}(\k,\e) $ and $3 \D
Q_{\nu}(\k,\e)=\rho_{\nu}(\e) \D_{\nu} (\k,\e)$, and in the limit $ k
\e_i \ll $ equation (\ref{eq:c7}) gives us 
\be
\d^{'}_{\nu}(\k,\e))+ \frac{2}{3} \mu^{'}(\k,\e)=0  \label{eq:i3}
\ee
and 
\be
\D^{'}_{\nu}(\k,\e))+ \frac{8}{15} \la^{'}(\k,\e)=0 \,. \label{eq:i4} 
\ee

We see that the evolution of both $\d_{\g}$ and $\d_{\nu}$ are governed by
the same equation and we can combine these two by defining 
\be
\d=\frac{\om_{\g} \d_{\g}+(n_{\nu}+1) \om_{\nu} \d_{\nu} }{\om_{\g} +
(n_{\nu}+1) \om_{\nu}} \,. \label{eq:i5}
\ee
The photon-baryon fluid has no anisotropic stresses and  there is no
contribution from the photons to $\D Q$.  

We also have equations (\ref{eq:a15}) and  (\ref{eq:a16}), which  can
now be written as   
\be
\mu^{''}+\frac{1}{\e} \mu^{'} =- \frac{6 \d}{\e^2} \label{eq:i6}
\ee
and  
\be
\la^{''}+\frac{1}{\e}(2 \la^{'} + \mu^{'}) = \frac{3}{\e^2}( \d + r_{\nu}
\D_{\nu})\,. \label{eq:i7}  
\ee

We simultaneously solve equations(\ref{eq:i2}), (\ref{eq:i3}),
(\ref{eq:i4}),  (\ref{eq:i6}) and  (\ref{eq:i7}) to obtain the
analytic form of the growing mode in the initial epoch, and the
solution for the metric perturbation can be written as
\be
 \mu(\k,\e)=\frac{\e^2 C}{2} \hspace{1cm} {\rm and} \hspace{1cm}
\la(\k,\e)= \frac{-10}{15+4 r_{\nu}} \mu(\k,\e)  \label{eq:18}
\ee
where $C$ is a constant which determines the amplitude of the
perturbation at the initial epoch. Note that we solve for $\mu$ and
$\la$ only up to an additive constant, and since we only encounter the
derivatives of $\mu$ and $\la$, the additive constant can be ignored
for our purposes. 

Using (\ref{eq:18}) we can write the solution for the perturbation in
the photon-baryon fluid as  
\be
\d_{\g}(\k,\e)=-\frac{\e^2 C}{3} \hspace{1cm} {\rm and} \hspace{1cm}
{\tilde \theta_{\g}}(\k,\e)=-\frac{\e^3 k^2 C}{36} \,. \label{eq:19}
\ee

and for each neutrino species we can write the distribution function as
\be
F_{\nu}(\k,\q,\e)= \frac{q}{6} \lsb \mu(\k,\e) + (1- 3 \a^2)
\la(\k,\e) \rsb \dq \fb_{\nu} \,. \label{eq:110}
\ee

The CDM particles do not contribute to the dynamics in the initial
epoch, and they move like test particles to which our synchronous
coordinate system is attached. We use equation (\ref{eq:x1}) to obtain
\be
\d_{\cdm}(\k,\e)=- \frac{1}{2} \mu(\k,\e)  \label{eq:111}
\ee
for the perturbation in the CDM component.  

We use these solutions to fix the initial conditions at the epoch
$\e_i$. We also assume that there is no significant decay of the
massive neutrino prior to the epoch $\e_i$ and we set the initial
density of the decay product to zero.  

\section{CMBR anisotropies due to the Sachs-Wolfe effect.}
At angular scales greater than a degree the dominant contribution to
the anisotropies in the CMBR is largely due to the Sachs-Wolfe
effect where the fluctuations in the CMBR temperature along any line
of sight  can be related to the derivative of the metric perturbation
$h^{'}_{a,b}(\x,\e)$  integrated along the photons trajectory form the
last scattering surface to the observer. For an observer located at
$\x_0$, the fluctuation in the CMBR temperature observed in the
direction $\n$ is given by   
\be
\frac{\Delta T}{T}(\n) =
 -{1 \over 2} \int_{\eta_{\rm dec}}^ {\eta_0}
 h^{'}_{ab} (\x_0-\n \ep, \ep) n^a n^b d \ep.
\label{eq:sw1}
\ee
where $\e_{\rm dec}$ refers to the value of the conformal time at the
epoch when the baryons and photons decouple, and $\e_0$ is the present
value of the conformal time. 

The angular dependence of this temperature fluctuation can be expanded
in terms of spherical harmonics $Y^m_l(\n)$
\be
\frac{\Delta T}{T}(\n) =\sum_{l,m} a^m_l Y^m_l(\n) \,.\label{eq:sw2}
\ee 
and the ensemble average of the square of the  expansion coefficients
gives the angular power spectrum 
\be
C_l=<\mid a^m_l \mid ^2>\, \label{eq:sw3}
\ee 
which, because the ensemble is statistically isotropic,  has no $m$
dependence. Writing the metric perturbation in terms of the Fourier
expansion and using equation (\ref{eq:sw1}) we write the CMBR angular
power spectrum as 
\be
C_l=\frac{1}{4 \pi} \int \frac{d^3 k}{(2 \pi)^3} \mid \D_l(\k) \mid^2
\label{eq:sw4} 
\ee
where 
\be
\D_l(\k,\e)=-2 \pi \int^1_{-1} d \a  P_l(\a) \int_{\eta_{\rm dec}}^
{\eta_0} \frac{1}{6} \lsb \mp(\k,\ep) e^{i \a k (\e_0 -\ep)}   + (1-3
\a^2) \lp(\k,\ep) e^{i \a k (\e_0 -\ep)} \rsb d \ep \,. \label{eq:sw5}   
\ee
which, using equations  (\ref{eq:c14}) and
(\ref{eq:c15}),  can be written as
\be
\D_l(\k,\e)=-(-i)^l \frac{2 \pi}{3}
\int_{\eta_{\rm dec}}^ {\eta_0}  \lsb \mp(\k,\ep) j_l( k (\e_0-\ep))
+  \lp(\k,\ep) \F_l(k (\e_0 -\ep)) \rsb d \ep \,. \label{eq:sw6}  
\ee  
where  $\F_l(k (\e_0 -\ep)$ is defined in equation (\ref{eq:c15}). We
use this in equation (\ref{eq:sw4})  to calculate the rms  quadrupole
$Q_{rms}=\sqrt{5 C_2/4 \pi}$ which we use to normalize the power
spectrum.   

\section{The  Numerical  Scheme and its Accuracy}
        To follow the evolution of the background universe we have
numerically solved equation (\ref{eq:a3}) for $a(\e)$,
along with equation (\ref{eq:d3}) for $\psi(q,\e)$ for a set of values
of $q$, and equation (\ref{eq:p3}) for $\om_R(\e)$.  The $q$ values
have been chosen so that they are appropriate for the Gauss-Laguerre
quadrature scheme, and we have used this method to evaluate the $q$
integral required to evaluate $\om_d(\e)$ (equation( \ref{eq:d5a})) and
the right hand side of equation (\ref{eq:p3}) at each time step. 
We have used 10 points to evaluate the $q$ integrals and we find that
there is no significant improvement if we increase the number of
points. 
We have used an adaptive step-size fifth order Runge-Kutta subroutine
`odeint'  (Press et. al. 1992) to follow the time evolution of  the
set of coupled ordinary differential equations.          
Along with the above mentioned quantities we   have also evolved the
quantities $\t(q,\e)$ and $\dq \psi(q,\e)$ which are required to
follow the perturbations. The intermediate values of all these
quantities are stored. We fit them with a cubic spline and use the
intermediate values in studying the evolution of the perturbation. 

        The time-steps have been chosen so as to achieve a relative
accuracy of $10^{-4}$. For all the models the background universe is
first evolved using $\Om_{\cdm0}=1$. It sometimes happens that present 
contribution from the decay products is quite large and the total
$\Om_0$ becomes greater than  one. In such cases we reduce the value of
$\Om_{\cdm0}$ and evolve the background again, and we keep on iterating
the process until the value of $\Om_0$ converges to within $1 \pm
0.001$. 

For the metric perturbation we numerically solve the two second order
differential equations (\ref{eq:a15}) and (\ref{eq:a16}) by converting
them into four first order equations. These equations are solved
together with equation (\ref{eq:x1}) for the CDM perturbations, and
equations (\ref{eq:x2}) and (\ref{eq:x3}) for the perturbations in the
photon-baryon fluid. This system of 7 differential equations is
evolved using `odeint' and the intermediate values of $\mp$ and $\lp$
at each time step are recorded and we fit these by a set of
overlapping cubic polynomials and these are used in following the
perturbations in the neutrinos. 
The metric perturbations are
coupled to the perturbations in the neutrinos, and these have to be
evaluated separately at each time step. For the massless neutrinos we
numerically evaluate the integrals in equations (\ref{eq:c16}) and
(\ref{eq:c17}) for every time step in `odeint'. Similarly, for the
massive neutrinos we numerically evaluate the $\ep$ integrals in
equations (\ref{eq:d8}), (\ref{eq:d9}), (\ref{eq:d10}) and
(\ref{eq:d11}) for a set of values of $q$ and we do the $q$ integrals
by a Gauss-Laguerre  quadrature. For the decay products we have
numerically evaluated the integrals in equations (\ref{eq:p6}) and
(\ref{eq:p7}) at every time-step in `odeint'.   

We have used 10 points for the  $q$ integrals, and we find that
increasing the number of points does not significantly change the
results for the feasible models in the range of $k$ that we have
considered. 

The results of the computations described above  yield the matter
transfer function  $T(k) \propto \mid \delta(\k,\e_0) \mid^2$. We
multiply this with $k$---the primordial Harrison Zel'dovich spectrum
to obtain the power spectrum.  This is normalized using 
equations (\ref{eq:sw4}) and (\ref{eq:sw6}) which we use to calculate
$Q_{rms}$. 

In our analysis we have ignored the baryon density and we use
$\Omega_B = 0$. In  addition, when following the  evolution of the
dark matter perturbations we have treated the photons as being tightly
coupled to baryons until the present epoch. Finally, we have
altogether ignored the interaction of the  photons with the electrons
and baryons when calculating the CMBR anisotropy.  
All these assumptions cause a few percent  error in our
results. These effects have been studied in detail for several  models 
(see for instance Bond 1996), and it is found  that the effect of
baryons can be included by scaling the value of $\Gamma$ as   
$\Gamma \times \exp(-2 \Omega_B)$. Primordial nucleosynthesis imposes
the restriction $\Omega_B h^2 \simeq 0.01$, and for 
$h = 0.5$ one expects a  $8 \%$ error if the baryons are left out. 
 
Treating photons as tightly coupled to baryons causes
the  sub-horizon scales  perturbations in photons to continue
to oscillate with  the same amplitude even after recombination. 
In reality, once the photons decouple from the baryons their  mean
free path becomes comparable to the size of the horizon and sub-horizon
perturbations in the photons start getting  wiped out as a result of
the free-streaming. However, the photons decouple in the matter
dominated era where they play no role in the dynamics of the dark
matter perturbations, and these effects are negligible.

 To check the accuracy of our numerical code we have compared our 
transfer function  for several models
with the runs of COSMICS (Bertschinger and Bode 1995) in the limit 
$\Omega_B \rightarrow 0$. For CDM models we get an agreement to better
than 5\% for $k \simlt 1 \, \Mi$. A comparison of the transfer functions
for various HDM and HCDM models allows us to test the reliability of
our  treatment of the massive neutrinos and we find that for the HDM
model, in  the range
$k< 3 \, \Mi$ our transfer function differs from the transfer
function calculated using COSMICS by less than $2 \%$, which 
indicates  that much of the error in our analysis come from the assumption
of tight coupling. This results in  larger  errors in HCDM 
models, for which,  in the range $0.1 \, \Mi < k \le 1 \, \Mi$, 
the error is within $10 \%$. 

We have not pinpointed the exact cause of this discrepancy, but some
of the possible sources are discussed below. One possible source of
the error could be the fact that in dealing with the massive
neutrinos one has to consider the evolution of neutrinos
with different momentum separately and then integrate over the
momentum. We have used only 10 values of momentum and the integration
was done using a Gauss-Laguerre quadrature where a few of the points
come with very low weights. COSMICS performs this integration using
8th-order Newton-Cotes method with as many as 128 points. This is one
of the sources of the error and it may be possible to overcome this by
using some other quadrature scheme (eg. Bond \& Szalay, 1983) and by
using more points. We have tried doubling the number of points but it
does not make a very big difference in the results in the range $k <
0.5 \, \Mi$.  Another possible cause for difference  could be the fact
that we have done all numerical integrations at a relative accuracy of
 $10^{-4}$ as compared to  the relative accuracy of $10^{-8}$ in
COSMICS. A point to be noted is that COSMICS uses various moments of
the collisionless Boltzmann equation in order to follow the evolution
of the perturbation in the neutrinos, and this involves a truncation
which is externally  enforced. Our method does not involve such a
truncation as we use the analytic solution of the collisionless
Boltzmann equation which is based on the method of characteristics,
but it has an extra cost as we have to do an integration  over the
entire past at every time step.  Our treatment also has another added
advantage in that it involves only differential equations, and does not
involve any algebraic equations, and we would expect it to be more
stable compared to methods based on a combination of algebraic and
differential equations. A little experimentation with the initial
conditions shows that when they are set arbitrarily (i.e. a mixture of
the growing and decaying modes) the solution goes over to the growing
mode  as expected and the effects of the decaying mode die away,
showing that our numerical scheme is indeed stable and does not
dependent critically  upon any fine tuning of the initial conditions.

Finally we note that a  $10\%$  inaccuracy  is acceptable in
calculating the  matter transfer function as it is   well below the
observational uncertainties. More accurate computations of decaying
neutrino transfer functions may be required in the future as
observations become more accurate. Also, a more intensive  treatment  
of the massive neutrinos is required at smaller scales which will be
essential to address issues related to galaxy formation in the
decaying neutrino models.  A need for more accurate computing will
also arise when computing CMBR anisotropies for comparing with
proposed future observations at small angular scales. Work is
currently in progress at improving the present numerical scheme so as
to be able to address these questions.

\begin{figure}
\caption{The contribution to $\omega(\e)$ from the various components
is shown as a function of the mode  $k=\pi/(c \e)$ which enters the
horizon at the epoch $\e$.} 
\label{fig:omm}
\plotfiddle{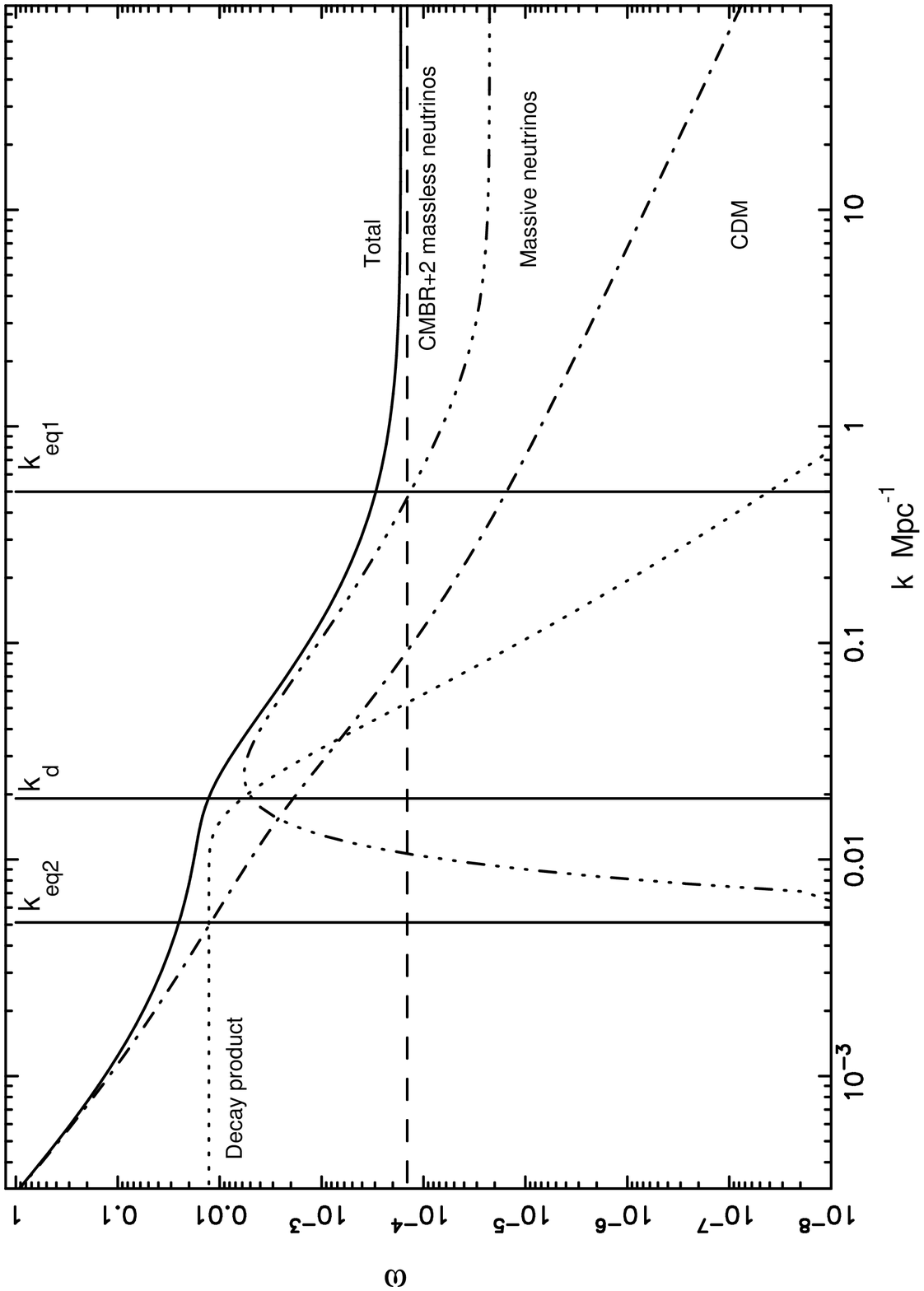}{3.in}{-90}{50}{50}{-200}{260}
\end{figure}

\begin{figure}
\caption{The solid curves show the  transfer
function for a decaying neutrino model with $h=0.5$, $m_{\nu}=200 \, \rm
eV$ and $t_d=10^{13}\, \rm s$. The smooth curve is the result of the
numerical computation whereas the jagged curve is based on the crude
approximation to the transfer function discussed in section 2. The
dashed curve shows the transfer function for the
$h=0.5$ CDM model.}
\label{fig:tk}  
\plotfiddle{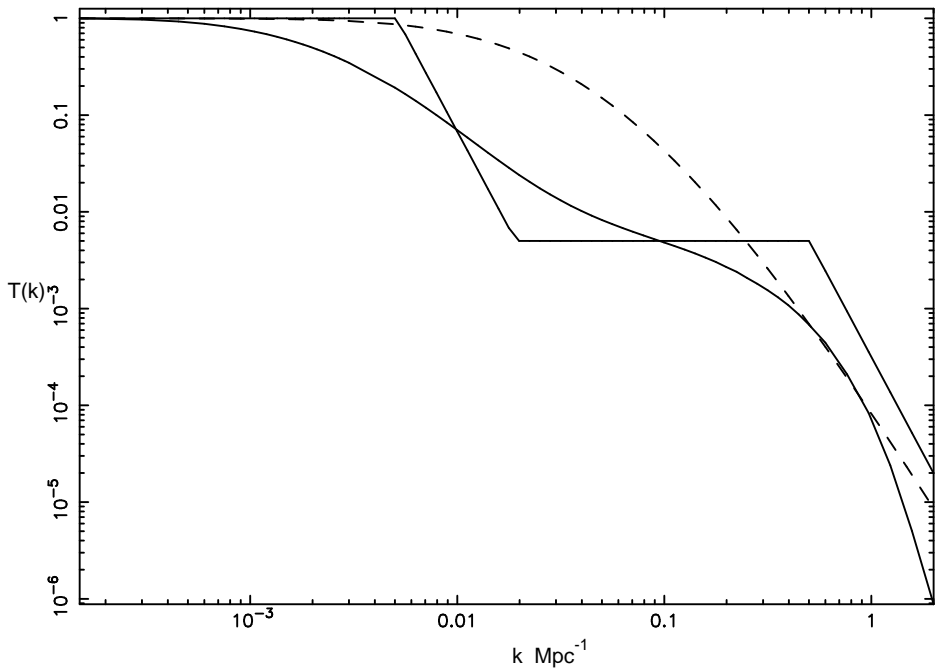}{3.in}{-90}{50}{50}{-200}{260}
\end{figure}

\begin{figure}
\caption{This shows the  power spectra for several decaying neutrino
models  all obtained by varying $m$ and $t_d$ keeping  $m_\nu^2 t_d$ a
constant.  The CDM power spectrum is shown for comparison. Also 
shown is the APM power spectrum (filled squares) and the best fit to
the LCRS power spectrum (dashed-dotted thick curve). 
$b =1.2$ has been assumed in plotting the APM and LCRS power
spectra and  $h = 0.5$ is used for all the power spectra shown here.} 
\label{fig:all}
\plotfiddle{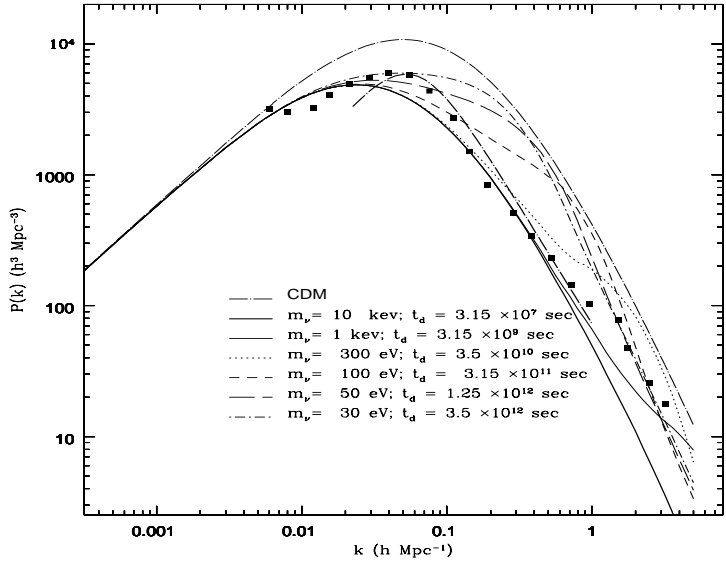}{2.8in}{0}{50}{40}{-180}{-80}
\end{figure}

\begin{figure}
\caption{ This shows how $\sigma_R$ ($R$ in $h^{-1} \, \rm Mpc$) changes
as a function of $m_\nu$ for a class of models for which  $m_\nu^2 ({\rm
keV})  t_d ({\rm yr}) = 100$.  The corresponding values for the 
standard CDM model are: $\sigma_8 =1.26  $,  $\sigma_4 = 2.32$,
$\sigma_1 = 3.81$, 
$\sigma_{0.5} = 5.75 $, and $\sigma_{0.25} = 10.89$. }
\label{fig:sig}
\plotfiddle{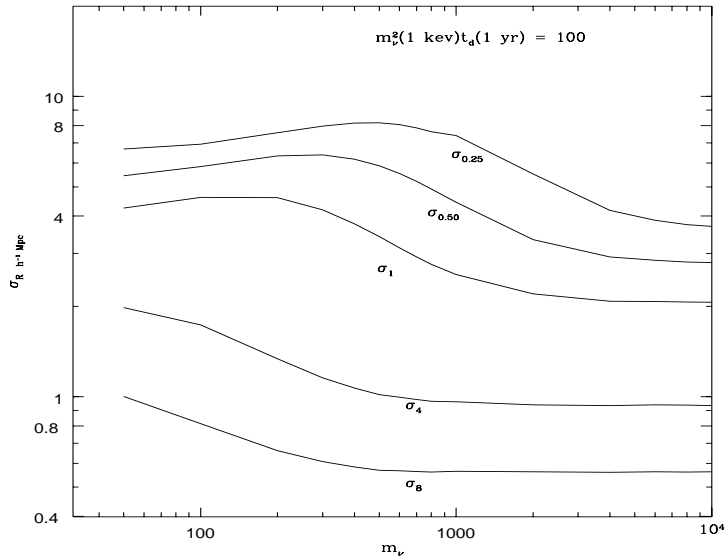}{2.8in}{0}{50}{40}{-180}{-80}
\end{figure}

\begin{figure}
\caption{This shows the power spectrum  for several models which give
acceptable value of $\sigma_8$ (Table~1). Most of these models are
ruled out on comparison 
to the APM and LCRS power spectra which have been plotted here for  $b
= 1.2$. These models also fail to make reasonable predictions for the
peculiar velocities (Table~1). h=.5 has been used here.  }
\label{fig:nal}
\plotfiddle{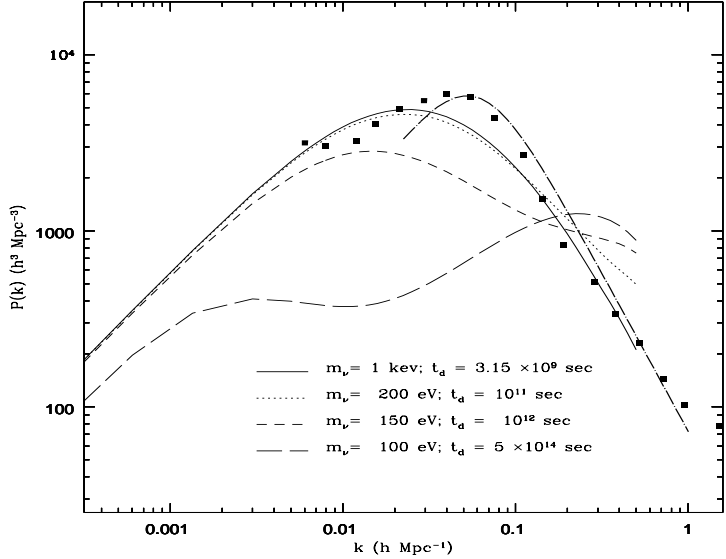}{2.8in}{0}{50}{40}{-180}{-80}
\end{figure}

\begin{figure}
\caption{Here we show the {\it r.m.s.} bulk velocity in spheres of radius $R$
after smoothing with a Gaussian of $12 h^{-1}$ Mpc. This is plotted as
a function of $R$ for the following decaying neutrino models, all of
which use $h=.5$:  (1) $m_\nu = 1 \, \rm keV$, $t_d = 3.15 \times 10^{9} \, 
\rm sec$ ({\em solid curve}), 
(2) $m_\nu = 200 \, \rm eV$, $t_d = 10^{11} \, \rm sec$ 
({\em dotted curve}), (3) $m_\nu = 150 \, \rm eV$, $t_d = 10^{12} \, \rm sec$
({\em short dashed curve}), and (4) $m_\nu = 100 \, \rm eV$, $t_d = 
5 \times 10^{14} \, \rm sec$ ({\em long dashed curve}).
The data points shown, $V_{40} = 388 \pm 67 \, \rm km/sec$ and $V_{60}
= 327 \pm 88$,  are taken from Bertschinger et al. (1990)}
\label{fig:vel}
\plotfiddle{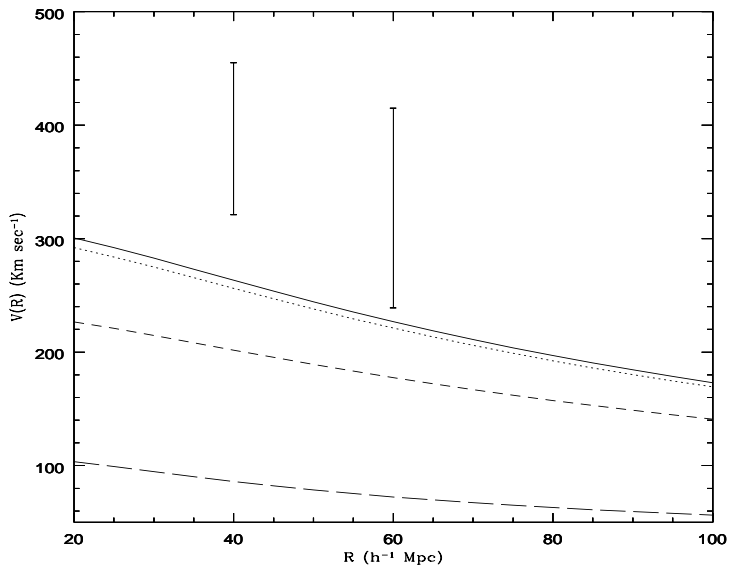}{2.8in}{0}{50}{40}{-180}{-80}
\end{figure}

\begin{figure}
\caption{Here we show the power spectrum for some of the radiatively
decaying neutrino models  (where a small fraction of neutrinos $B
\simeq 10^{-5}\hbox{--}10^{-7}$ decay into photons) which  satisfy 
various observations of the  ionization of the intergalactic medium
 at high redshifts.}
\label{fig:dec}
\plotfiddle{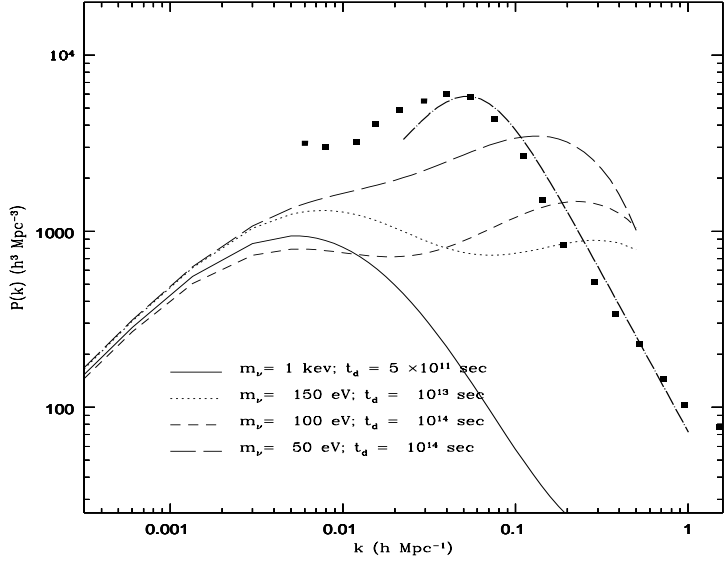}{2.8in}{0}{50}{40}{-180}{-80}
\end{figure}
\begin{figure}
\caption{The parameter space allowed by structure formation bounds 
({\em shaded region}) corresponding to  $0.2 \le \Gamma
\le 0.3$  where  $\Gamma(m_\nu,t_d)$ is given by Eq.~(\ref{eq:r1}) 
is shown along with the region on the $m_\nu\hbox{--}t_d$ plane
allowed by IGM considerations (i.e. Eqs. (\ref{eq:ra}) and (\ref{eq:ra1}))
({\em hatched region}).} 
\label{fig:reg}
\plotfiddle{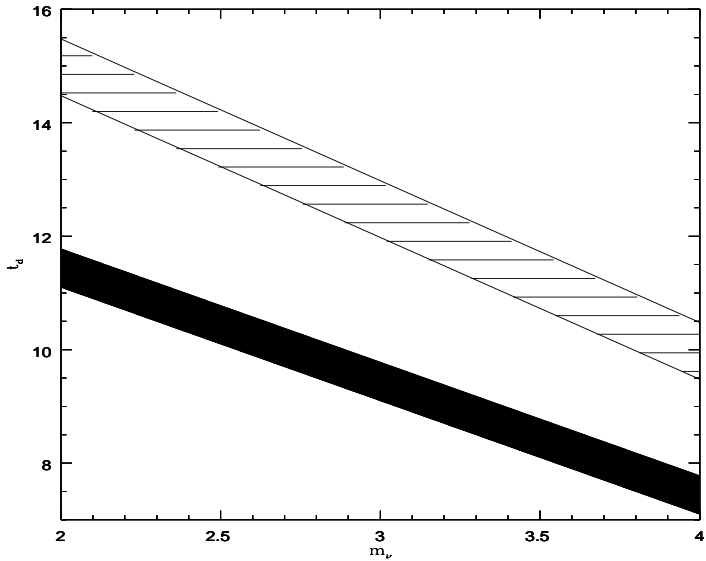}{2.8in}{0}{50}{40}{-180}{-80}
\end{figure}

\begin{figure}
\caption {The density of collapsed baryons $\Omega_{\rm col}$ for
several  structure formation models is compared  to the 
density of  neutral hydrogen $\Omega_{\rm HI}$ observed in the  damped
Lyman-$\alpha$ clouds. The shaded region corresponds to Eq.~(\ref{eq:hyd})
( Lanzetta, Wolfe, and Turnshek  1995). The data point at $z \simeq 4$
is taken from Storrie-Lombardi et al. (1995). }
\label{fig:dam}
\plotfiddle{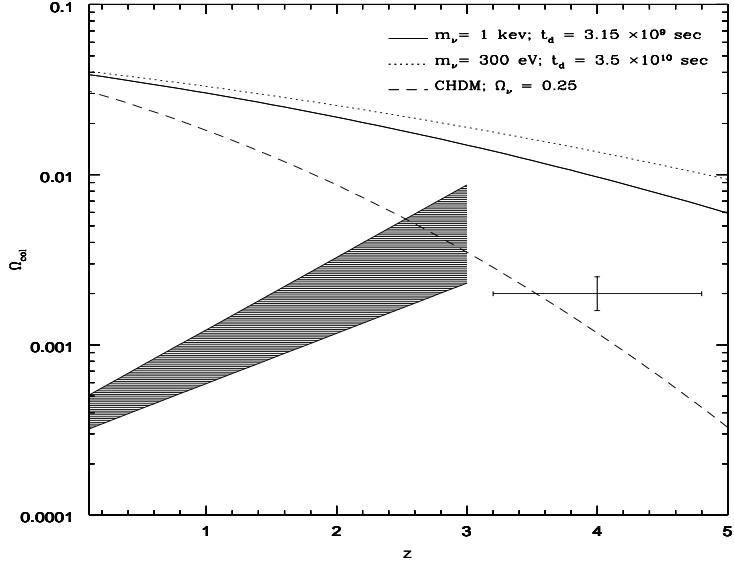}{2.8in}{0}{50}{40}{-180}{-80}
\end{figure}

\begin{table}
\center{Table 1: Values of $\sigma_8$ and {\it r.m.s.} bulk velocity 
in km/s in spheres of radius $40 h^{-1} \, \rm Mpc$ and $60 h^{-1} \,
\rm Mpc$ after smoothing with a Gaussian of $12 h^{-1}$ Mpc.} 
\vskip 0.2in
\centerline{\begin{tabular}{lcccc}
\tableline
\tableline
{$m_\nu$ (eV)} & {$t_d$ (sec) }&{$\sigma_8$}&{$V_{40}$}&{$V_{60}$}\\
\tableline
$10^{4}$ & $3.15 \times 10^{7}$ &0.55& 263&226\\
$10^{3}$ & $3.15 \times 10^{7}$ &1.13& 345&286\\
$10^{3}$ & $3.15 \times 10^{9}$ &0.56& 263&226\\ 
$10^{3}$ & $5 \times 10^{11}$ &0.1& 95&88\\ 
$500$ & $1.26 \times 10^{10}$ &0.57& 262&226\\
$300$ & $3.5 \times 10^{10}$ &0.6& 263&226\\
$150$ & $1.4 \times 10^{11}$ &0.71& 264&226\\
$150$ & $ 10^{12}$ &0.62& 201&177\\
$150$ & $ 10^{13}$ &0.58& 132&118\\
$100$ & $3.15\times 10^{11}$ &0.81& 266&228\\
$100$ & $10^{14}$ &0.72& 113&98\\
$100$ & $5\times 10^{14}$ &0.69& 86&72\\
$50$ & $1.25\times 10^{12}$ &1.0& 274&232\\
$50$ & $ 10^{14}$ &0.98&182&153\\
$30$ & $3.5\times 10^{12}$ &1.08& 285&239\\
\tableline
\tableline
\end{tabular}}
\end{table}

\end{document}